\documentclass[aps, prd, showpacs, superscriptaddress, preprintnumbers, nofootinbib, twocolumn]{revtex4-1}
\usepackage[pdftex]{graphicx, color}
\usepackage{amsmath, amssymb, amscd, latexsym, bm, braket}
\usepackage[colorlinks=true,pdfstartview=FitV,linkcolor=blue,citecolor=magenta,urlcolor=blue,bookmarks=true,bookmarksnumbered=true]{hyperref}

\newcommand{\Slash}[1]{{\ooalign{\hfil/\hfil\crcr$#1$}}}

\begin{document}
\title{Dynamically assisted Schwinger mechanism and chirality production in parallel electromagnetic field}

\author{Hidetoshi Taya}
\email{h\_taya@fudan.edu.cn}
\address{Department of Physics and Center for Field Theory and Particle Physics, Fudan University, Shanghai, 200433, China }

\date{\today}

\begin{abstract}

We study particle and chirality production from the vacuum in the presence of a slow strong parallel electromagnetic field superimposed by a fast weak perturbative electromagnetic field.  We derive an analytical formula for the particle and chirality production number based on the perturbation theory in the Furry picture.  With the formula, we analytically discuss the interplay and dynamical assistance between the Schwinger mechanism and one-photon pair production and clarify effects of a parallel slow strong magnetic field.  We also show that the dynamical assistance can significantly enhance chirality production, and that a sizable amount of chirality can be produced even for massive particles.  Phenomenological applications including heavy-ion collisions and intense laser experiments are also discussed.  

\end{abstract}

\maketitle

\section{Introduction}

\begin{figure}[!t]
\begin{center}
\includegraphics[width=0.4\textwidth]{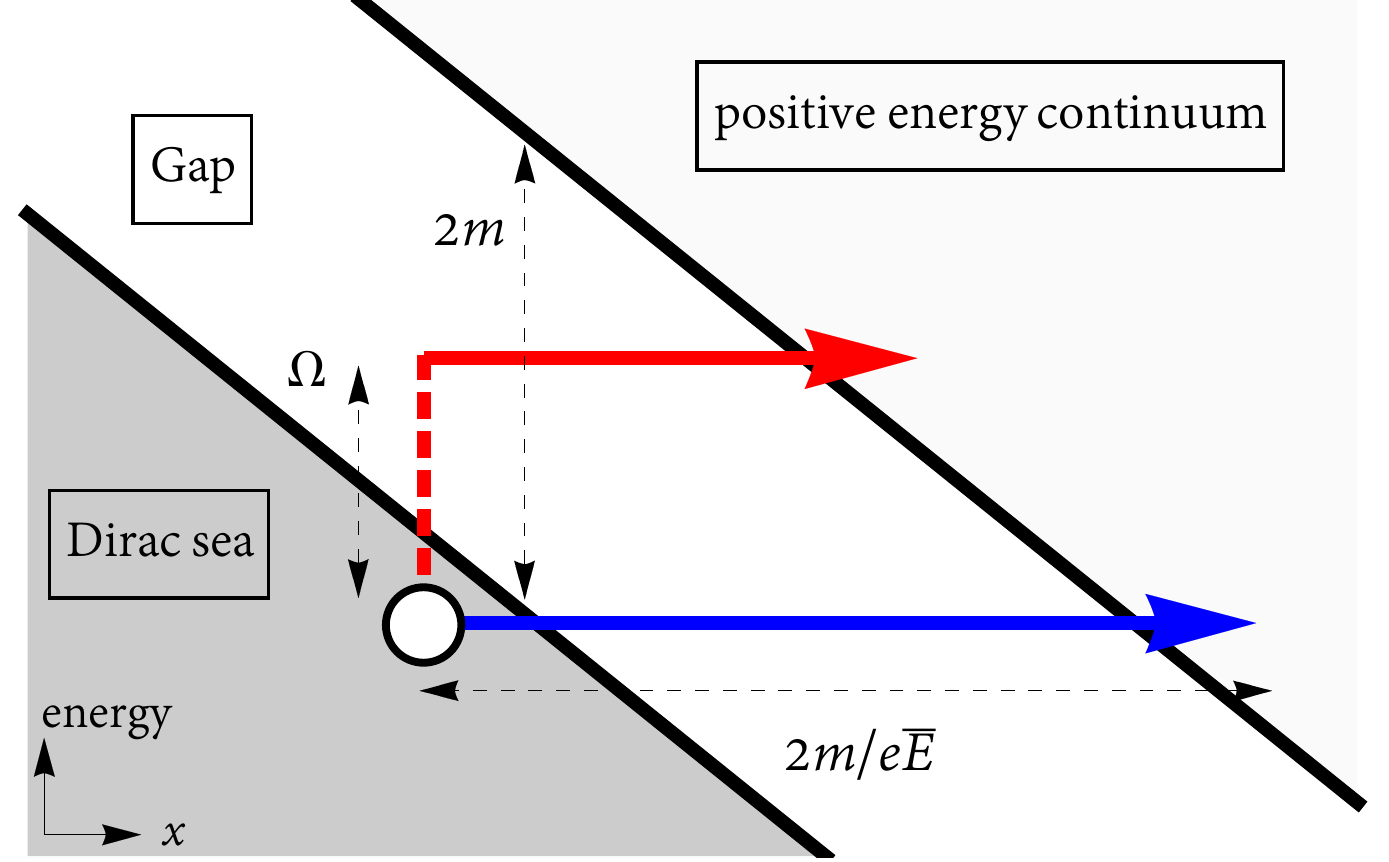}
\caption{\label{fig-1} Band structure of the QED vacuum in the presence of a slow strong electric field applied in the $x$-direction, $V(x) = - e\bar{E} x$, and schematic pictures of the Schwinger mechanism (blue arrow) and the dynamically assisted Schwinger mechanism (red arrow).  $2m$, with $m$ being mass, is the gap size, and $\Omega$ represents energy supplied by a fast weak electromagnetic field in the dynamically assisted Schwinger mechanism.  }
\end{center}
\end{figure}

One of the most remarkable predictions of quantum electrodynamics (QED) is spontaneous particle production from the vacuum in the presence of a slow strong electromagnetic field (for review, see Refs.~\cite{gel15, ruf10, dun05}).  This remarkable prediction was first made by Sauter in 1931 \cite{sau31}.  Sauter's idea was sophisticated later by Heisenberg and Euler \cite{hei36} and by Schwinger \cite{sch51}, who fully formulated the idea within quantum field theory for the first time.  Thus, the vacuum particle production by a slow strong electromagnetic field is called {\it the Schwinger mechanism}.  The Schwinger mechanism is essentially an electric effect.  In the presence of a slow strong electric field, there occurs a level crossing between the Dirac sea and the positive energy continuum (see Fig.~\ref{fig-1}).  Then, an electron filling the Dirac sea can tunnel into the positive energy continuum, leaving a hole in the Dirac sea.  Thus, a pair of an electron and a positron is spontaneously produced, which can be understood as a QED analog of electrical breakdown or the Landau-Zener transition in materials \cite{lan32, zen32, stu32, maj32}.

The production number of the Schwinger mechanism is determined by the tunneling rate, which decreases with increasing the gap size $\sim m$ and the tunneling length $\sim m/e\bar{E}$ (see Fig.~\ref{fig-1}), where $m$, $e>0$, and $\bar{E}$ are mass, the QED coupling constant, and the electric field strength, respectively.  Therefore, one may expect that the production number is suppressed by an exponential of $m \times m/e\bar{E}$.  Indeed, Schwinger \cite{sch51} showed that the production number of electrons $N$ and positrons $\bar{N}$ for a constant and homogeneous electric field is given by
\begin{align}
	N = \bar{N} = VT \times \frac{(e\bar{E})^2}{4\pi^3} {\rm e}^{-\pi \frac{m^2}{e\bar{E}} } \label{eq1}
\end{align}
with $V$ and $T$ being the spatial volume and the whole time-interval, respectively.  For a more general electro{\it magnetic} field configuration, the production number formula reads \cite{sch51, nik70, mar72, gav96, tan09, hid11a, hid11b, hat20} (see also Refs.~\cite{pop72, iwa09, tan10, tan12, kar19a, kar19b, she19} for more discussions on magnetic-field effects)
\begin{align}
	N = \bar{N} = VT \times \frac{(e\bar{E})^2}{4\pi^3} {\rm e}^{ -\pi \frac{m^2}{e\bar{E}} } \times  \pi \frac{e\bar{B}}{e\bar{E}} \coth \left[ \pi \frac{e\bar{B}}{e\bar{E}} \right] . \label{eq2}
\end{align}
Remark that it is sufficient to consider a parallel electromagnetic field configuration $\bar{\bm E} \parallel \bar{\bm B}$, which can cover all the possible values of the Lorentz invariants ${\mathcal F} \equiv \bar{\bm E}^2 - \bar{\bm B}^2$ and ${\mathcal G} \equiv \bar{\bm E}\cdot\bar{\bm B}$.  Equation~(\ref{eq2}) shows that the addition of a parallel magnetic field enhances the production number  by the factor $\pi (e\bar{B}/e\bar{E}) \coth \left[ \pi (e\bar{B}/e\bar{E}) \right] > 1 $\footnote{Precisely speaking, a parallel magnetic field enhances the production number {\it except} for scalar particles, whose lowest energy level increases with $e\bar{B}$ as $\sqrt{m^2+|e\bar{B}|}$ and hence its production is exponentially suppressed \cite{pop72, tan09}.  The production of vector particles (e.g., gluon and $W$-boson) or, more generally, higher spin particles, is enhanced more strongly than that of spinor particles \cite{mar72, tan12} because of the existence of unstable modes (Nielsen-Olesen instability \cite{nie78a, nie78b, cha79}).  }.  The Landau quantization is the essence of this enhancement.  Namely, the Landau quantization discretizes the transverse mass $\sqrt{m^2 + {\bm p}_{\perp}^2} \to \sqrt{m^2+|e\bar{B}|(2n+1-s)}$ (where $n = 0, 1, 2, \ldots \in {\mathbb N}$ labels the Landau level, and $s=\pm 1$ is the spin component with respect to the magnetic field direction) and, accordingly, the phase-space $\int {\rm d}^2{\bm p}_{\perp} \to e\bar{B} \sum_n$.  The phase-space linearly increases with $e\bar{B}$, which essentially accounts for the enhancement of the production number of electrons and positrons.  The mass suppression factor $\exp \left[ -\pi m^2/ e\bar{E} \right]$ remains the same since the gap size $\sim m$ and the tunneling length $\sim m/e\bar{E}$ are unchanged by the Landau quantization, or the presence of a magnetic field, for spinor particles.

The Schwinger mechanism (or the idea of the Schwinger mechanism) has a wide spectrum of phenomenological applications including, to name a few, the early Universe \cite{kob14, sor19, sob18, sha17, sta18, kob19, dom20}, condensed matter/materials \cite{oka08, mat19, gre05, all08, kas12, szp12, fra15, kas17, ali19}, and axion production \cite{gri99, gri02, hua20}.  In particular, application to heavy-ion collisions has received much attention over the decades.  Just after a collision of heavy ions at the Relativistic Heavy Ion Collider (RHIC) and the Large Hadron Collider (LHC), a strong (chromo-)electromagnetic field, which is sometimes called {\it glamsa}, is produced \cite{low75, nus75, kov95a, kov95b, lap06}.  It is widely recognized that particle production from the glasma via the Schwinger mechanism is the essence of the formation of the quark-gluon plasma in heavy-ion collisions \cite{low75, nus75, gle83, kaj85, gat87, tan09, tay17}, while its detailed understanding is still lacking.  In addition, the glasma has a parallel (chromo-)electromagnetic field configuration, and therefore produces chirality through the chiral anomaly \cite{adl69, bel69, nie83}.  The chirality imbalance may induce novel anomalous transport phenomena (see Refs.~\cite{kha14, hua16} for review) such as the chiral magnetic effect \cite{fuk10, kha08, fuk08}.  Experimental search for anomalous transport phenomena is a very active topic in heavy-ion collision experiments; see Ref.~\cite{zha19} for a recent review.  Therefore, it is important to deepen our understanding of particle and chirality production in the presence of a parallel strong electromagnetic field via the Schwinger mechanism and discuss possible observable consequences in, e.g., heavy-ion collisions.

The Schwinger mechanism has not been observed in laboratory experiments yet, despite its great theoretical/phenomenological interest.  The most promising way to observe the Schwinger mechanism in laboratory experiments is to use intense lasers.  However, the Schwinger mechanism is exponentially suppressed by the mass factor as in Eqs.~(\ref{eq1}) and (\ref{eq2}), and hence an electric field strength of the order of the electron mass scale $e \bar{E}_{\rm cr} \equiv m_e^2 \sim \sqrt{10^{29}\;{\rm W/cm^2}}$ is required for the Schwinger mechanism to be manifest.  The available field strength at the present is limited to $e\bar{E} \sim \sqrt{10^{22}\;{\rm W/cm^2}}$ \cite{yan08}.  Up-coming intense laser facilities such as the Extreme Light Infrastructure (ELI) and at the Exawatt Center for Extreme Light Studies (XCELS) are expected to reach $e\bar{E} \sim \sqrt{ 10^{24{\rm -}25}\;{\rm W/cm^2}}$ but are still weaker than the critical field strength $\bar{E}_{\rm cr}$ by several orders of the magnitude (see Ref.~\cite{pia12} for a review of the current experimental situation).  Therefore, it is still difficult within the current laser technology to directly observe the Schwinger mechanism.

The experimental difficulty motivated theorists to investigate how to enhance the Schwinger mechanism to observe it with a {\it subcritical} electromagnetic field.  One of the promising ideas is {\it the dynamically assisted Schwinger mechanism} \cite{sch08, piz09, dun09, mon10a, mon10b}, which is an analog of the Franz-Keldysh effect in semi-conductor physics \cite{fra58, kel58, fk, fk2}.  The idea is to superimpose a weak fast electromagnetic field with a large Keldysh parameter \cite{tay14, bre70, pop72, kel65} (or inject an energetic dynamical photon) onto a slow strong electromagnetic field (see Fig.~\ref{fig-1}).  Then, the weak fact electromagnetic field perturbatively interacts with electrons in the Dirac sea and supplies energy, which is depicted by the red dashed line in Fig.~\ref{fig-1}.  The electrons come out from the Dirac sea, and the gap size as well as the tunneling length are effectively reduced, which implies a reduction of the mass suppression factor in the production number formulas (\ref{eq1}) and (\ref{eq2}).  Therefore, the Schwinger mechanism is enhanced, and the enhancement was found to be significant \cite{sch08, piz09, dun09, mon10a, mon10b}, providing a hope to observe the Schwinger mechanism, albeit indirectly, with intense lasers in the near future.

A number of aspects of the dynamically assisted Schwinger mechanism have been investigated/clarified within, for example, the worldline instanton formalism \cite{dun05a, dun06a, dun06b}, numerical simulations based on the quantum kinetic theory \cite{smo97, sch98}, and the recently developed perturbation theory in the Furry picture \cite{fur51, fra81, fra91, gre17, gre18, gre19, fk, fk2}.  Examples include optimization of a field profile \cite{koh13, heb-14, abd13, lin15, fra17, don20}, momentum distribution \cite{heb09, ort11, chr12, lli14, pan15, dum10, dum11, fk, fk2, gre17, gre19}, finite size effects \cite{aba19, aba-19}, and spatially dependent perturbations \cite{cop16, sch16, ale17, ale18}.  In particular, it was clarified in Refs.~\cite{fk, fk2} that the dynamically assisted Schwinger mechanism can be understood in terms of the interplay between the Schwinger mechanism and one-photon pair production process $\gamma + \bar{E} \to e^+ e^-$.  Namely, in the presence of both a slow strong electromagnetic field and a fast weak perturbation, the Schwinger mechanism and one-photon pair production dominate the particle production if the frequencies (i.e., the energy supply) of the perturbation are small and large, respectively.  At intermediate frequencies, both production mechanisms take place and assist each other to significantly enhance the particle production; i.e., the dynamically assisted Schwinger mechanism occurs.

Although there are many preceding studies on the dynamically assisted Schwinger mechanism, they focus mostly on a slow strong purely {\it electric} field configuration and, therefore, effects of a parallel strong magnetic field are less understood (effects of a transverse strong magnetic field associated with a finite-sized strong electric field were recently discussed in Ref.~\cite{koh-19}).  In other words, the dynamically assisted Schwinger mechanism for a non-vanishing Lorentz invariant ${\mathcal G} \neq 0$ is unexplored, and it is unclear how the dynamical assistance affects chirality production.  These questions are important to complete our understanding of the dynamically assisted Schwinger mechanism as well as to discuss phenomenology such as anomalous transport phenomena in heavy-ion collisions and to propose new observables in the up-coming laser experiments.  To the best of our knowledge, Ref.~\cite{cop16} is the only work in this direction at the present.  Ref.~\cite{cop16} investigated the dynamically assisted Schwinger mechanism in the presence of a parallel slow strong electromagnetic field configuration within the worldline instanton formalism just for a slowly varying perturbation (only for which the worldline instanton formalism can be justified) with Sauter-type spacetime dependence, and chirality production was not discussed explicitly.  It is important to go beyond a slowly varying perturbation because the dynamically assisted Schwinger mechanism is, and accordingly chirality production is expected to be, enhanced more strongly for faster perturbations \cite{fk, fk2}.  Indeed, chirality production for a pulsed electric field was discussed in Ref.~\cite{amb83}, finding that chirality production (by massive particles) decreases with increasing duration of the electric field.  Since no chirality should be produced in the limit of vanishing duration, we can naturally expect that there exists an optimal duration or frequency for the dynamical assistance to maximize chirality production.

The purpose of the present paper is to discuss effects of a parallel slow strong magnetic field on the dynamically assisted Schwinger mechanism and its impacts on chirality production.  This shall be achieved in the following manner: In Sec.~\ref{sec2}, we discuss the number of particles produced from the vacuum by a slow strong parallel electromagnetic field superimposed by a fast weak perturbative electromagnetic field.  We derive an analytical formula for the production number by extending the perturbation theory in the Furry picture for a purely slow strong electric field configuration \cite{fur51, fra81, fra91, gre17, gre18, gre19, fk, fk2} to include a parallel magnetic component.  With this formula, we analytically discuss the interplay between the Schwinger mechanism and one-photon pair production and the dynamical assistance between the two production mechanisms in the presence of a parallel slow strong magnetic field.  Advantages of the perturbation theory in the Furry picture are that (i) the production number formula is applicable to perturbations with {\it arbitrary} time-dependence and hence it is valid even for a fast perturbation (the validity was explicitly tested by comparing with a numerical approach in Refs.~\cite{gre17, gre19, fk, fk2}), which is not accessible with approaches based on adiabatic approximations such as the worldline instanton formalism \cite{dun05a, dun06a, dun06b}; (ii) one can manifestly discuss the interplay and dynamical assistance between the Schwinger mechanism and one-photon pair production, which is not feasible within other approaches; and (iii) the formula can easily be applied to chirality production.  In Sec.~\ref{sec3}, we discuss chirality production for the same field configuration.  We first derive an analytical formula for chirality production by explicitly evaluating an in-in vacuum expectation value of the chirality operator and using the production number formula.  Based on the formula, we investigate how the dynamically assisted Schwinger mechanism affects chirality production.  In Sec.~\ref{sec4}, we summarize our findings and discuss future work and implications for or applications to, e.g., heavy-ion collisions and intense laser experiments.

\section{Production number} \label{sec2}

In this section, we discuss the number of particles produced from the vacuum in the presence of a slow strong electromagnetic field superimposed by a fast weak perturbative electromagnetic field.  Our formulation is based on the perturbation theory in the Furry picture \cite{fur51, fra81, fra91}, which was recently applied to a case where a slow strong electromagnetic field is purely electric \cite{gre17, gre18, gre19, fk, fk2}.  We generalize the preceding calculations by including a parallel magnetic component and derive an analytical formula for the production number (see Sec.~\ref{sec2a}).  Based on the formula, we analytically discuss the particle production and clarify how the existence of a slow strong parallel magnetic field affects the interplay between the Schwinger mechanism and one-photon pair production and the dynamical assistance between the two mechanisms (see Secs.~\ref{sec2b} and \ref{sec2c}).

\subsection{Perturbation theory in the Furry picture} \label{sec2a}

We explain the perturbation theory in the Furry picture by following Ref.~\cite{fk} and derive a formula for the number of particles produced from the vacuum in the presence of a slow strong electromagnetic field $\bar{A}_\mu$ superimposed by a fast weak perturbative electromagnetic field ${\mathcal A}_{\mu} \ll \bar{A}_\mu$.  In this work, for the sake of simplicity, we neglect back-reaction from produced particles and treat the electromagnetic fields just classically, i.e., higher order quantum processes such as bremsstrahlung are neglected.

We first solve the Dirac equation, 
\begin{align}
	\left[ {\rm i}\Slash{\partial} - e\bar{A} - m \right] \hat{\psi} = e \Slash{\mathcal A} \hat{\psi},   \label{eq-1}
\end{align}
perturbatively in terms of the perturbative field ${\mathcal A}_{\mu}$ while the interaction with the strong field $\bar{A}_\mu$ is treated non-perturbatively.  To do this, we introduce a (retarded) Green function $S_{\rm R}$ which is fully dressed by the strong field $\bar{A}_\mu$ as
\begin{align}
	\left\{ \begin{array}{l}
		\displaystyle \left[ {\rm i}\Slash{\partial}_x - e\bar{A}(x) - m \right] S_{\rm R} (x,y) = \delta^4(x-y) \vspace*{2mm}\\
		\displaystyle S_{\rm R} (x,y) = 0\ {\rm for}\ x^0 - y^0 < 0  
	\end{array} \right. .
\end{align}
With the Green function $S_{\rm R}$, we can write down a formal solution of the Dirac equation (\ref{eq-1}) as
\begin{align}
	\hat{\psi}(x) = \sqrt{Z}\hat{\psi}^{\rm in}(x) + e\int {\rm d}^4y\;S_{\rm R}(x,y) \Slash{\mathcal A}(y) \hat{\psi}(y). \label{eq-3}
\end{align}
Here, we assumed that the perturbative field ${\mathcal A}_{\mu}$ adiabatically goes off at the infinite past and future and required the Lehmann-Symanzik-Zimmermann (LSZ) asymptotic condition \cite{leh55}
\begin{align}
	\lim_{x^0 \to -\infty/+\infty} \hat{\psi}(x) = \sqrt{Z} \hat{\psi}^{\rm (in/out)} 
\end{align}
with $Z$ being the field renormalization constant and $\hat{\psi}^{\rm (in/out)}$ being the asymptotic field operator at $x^0 \to -\infty/+\infty$.  Notice that $\hat{\psi}^{\rm (in)} \neq \hat{\psi}^{\rm (out)}$ in general because of the interaction with the electromagnetic fields $\bar{A}_\mu$ and ${\mathcal A}_\mu$.  The asymptotic field operator $\hat{\psi}^{\rm as}$ (${\rm as}={\rm in},{\rm out}$) is the solution of the Dirac equation without the perturbative field ${\mathcal A}_{\mu}$ and can be expressed as a mode integral
\begin{align}
	\hat{\psi}^{\rm (as)}(x) = \sum_i \!\!\!\!\!\!\!\! \int \left[ {}_+ \psi^{\rm (as)}_{i}(x) \hat{a}^{\rm (as)}_{i} + {}_- \psi^{\rm (as)}_{i}(x) \hat{b}^{{\rm (as)}\dagger}_{i} \right] , \label{eq-5}
\end{align}
where $\displaystyle \sum_i \!\!\!\!\!\!\!\! \int$ denotes summation/integration over all the quantum numbers $i$ (e.g., momentum, spin) and the mode functions ${}_{+} \psi_{i}^{\rm (as)}$ and ${}_{-} \psi_{i}^{\rm (as)}$ are the two independent solutions of
\begin{align}
	0 = \left[ {\rm i}\Slash{\partial} - e\bar{\Slash{A}} - m \right]  {}_{\pm} \psi^{\rm (as)}_i \label{eq-6}
\end{align}
with normalization
\begin{align}
	\left\{\begin{array}{l}
		\displaystyle \delta_{i,i'} = \int {\rm d}^3{\bm x}\;{}_{\pm} \psi^{{\rm (as)}\dagger}_i {}_{\pm} \psi^{{\rm (as)}}_{i'} \vspace*{2mm}\\
		\displaystyle 0 = \int {\rm d}^3{\bm x}\;{}_{\mp} \psi^{{\rm (as)}\dagger}_i {}_{\pm} \psi^{{\rm (as)}}_{i'}
	\end{array}\right. .  \label{eq-7}
\end{align}
The subscript $\pm$ specifies the positive and negative frequency modes of the mode function ${}_{\pm} \psi_{i}^{\rm (as)}$.  Namely, we identify the positive and negative frequency mode functions at the asymptotic time, ${}_{+} \psi_{i}^{\rm (as)}$ and ${}_{-} \psi_{i}^{\rm (as)}$, respectively, by
\begin{align}
	\lim_{x^0 \to -\infty/+\infty} {}_{\pm} \psi^{{\rm (in/out)}\dagger}_i \propto \exp \left[ \mp {\rm i} \int^{x^0} {\rm d}{x'}^0\; \omega_i \right],
\end{align}
where $\omega_i>0$ is the one-particle energy at the asymptotic time.

After the canonical quantization procedure, we can interpret $\hat{a}^{\rm (as)}_{i}$ and $\hat{b}^{\rm (as)}_{i}$ as annihilation operators of a particle and an anti-particle at the corresponding asymptotic states, respectively.  For the normalization (\ref{eq-7}), the commutation relations for $\hat{a}^{\rm (as)}_{i}$ and $\hat{b}^{\rm (as)}_{i}$ read 
\begin{align}
	\left\{ \begin{array}{l}
		\displaystyle \delta_{i,i'} = \{ \hat{a}^{\rm (as)}_{i}, \hat{a}^{{\rm (as)}\dagger}_{i'} \} = \{ \hat{b}^{\rm (as)}_{i}, \hat{b}^{{\rm (as)}\dagger}_{i'} \} \vspace*{2mm}\\
		\displaystyle 0 = ({\rm others})
	\end{array}\right. .
\end{align}

The annihilation operators at the infinite past and future do not coincide with each other in the presence of the external electromagnetic fields $\bar{A}_\mu$ and ${\mathcal A}_\mu$.  Indeed, from the formal solution (\ref{eq-3}), we find
\begin{align}
	\!\!
	\begin{pmatrix}
		\hat{a}^{\rm (out)}_{i} \\ 
		\hat{b}^{{\rm (out)}\dagger}_{i} 
	\end{pmatrix}
	&\!=\!
	\int {\rm d}^3{\bm x} 
	\begin{pmatrix}
		{}_{+} \psi^{{\rm (out)}\dagger}_{i} \\
		{}_{-} \psi^{{\rm (out)}\dagger}_{i} 
	\end{pmatrix}
	\hat{\psi}^{\rm out} \nonumber\\
	&\!=\!
	\lim_{x^0 \to \infty}
	\int {\rm d}^3{\bm x} 
	\begin{pmatrix}
		{}_{+} \psi^{{\rm (out)}\dagger}_{i} \\
		{}_{-} \psi^{{\rm (out)}\dagger}_{i} 
	\end{pmatrix} \nonumber\\
	&\!\quad\!\times\! \left[
	\hat{\psi}^{\rm in}
	\!+\!
	e \!\int \!{\rm d}^4y\;S_{\rm R}(x,y) \Slash{\mathcal A}(y) \hat{\psi}^{\rm (in)}(y) \!+\! {\mathcal O}(e^2) \right] \nonumber\\
	&\!\!\equiv\! \sum_{k=0}^{\infty} e^k
			\begin{pmatrix}
				\hat{a}^{({\rm out};k)}_{i} \\ 
				\hat{b}^{({\rm out};k)\dagger}_{i} 
			\end{pmatrix}.
\end{align}
By using
\begin{align}
	\!\!\!S_{\rm R}(x,y) 
		&\!=\! -{\rm i} \theta(x^0-y^0) \{ \hat{\psi}^{\rm (as)}(x) , \hat{\bar{\psi}}^{\rm (as)}(y) \} \nonumber\\
		&\!=\! -{\rm i} \theta(x^0-y^0) \sum_i \!\!\!\!\!\!\!\! \int \nonumber\\
		&\!\quad\!\times\!\! \left[ {}_{+\!} \psi^{\rm (as)}_{i}\!(x){}_{+\!} \bar{\psi}^{\rm (as)}_{i}\!(y) \!+\! {}_{-\!} \psi^{\rm (as)}_{i}\!(x) {}_{-\!} \bar{\psi}^{\rm (as)}_{i}\!(y) \right]\!,\!\!\!
\end{align}
we can explicitly write down $\hat{a}^{({\rm out};k)}_{i}$ and $\hat{b}^{({\rm out};k)}_{i}$.  Up to $k \leq 1$, we find
\begin{subequations}
\label{eq-11}
\begin{align}
	\hat{a}^{({\rm out};0)}_{i} 
		&\!=\! \sum_{i'} \!\!\!\!\!\!\!\! \int \left[ \left( \int \!{\rm d}^3{\bm x} {}_{+} \psi^{{\rm (out)}\dagger}_{i} {}_+ \psi^{\rm (in)}_{i'} \!\right) \hat{a}^{\rm (in)}_{i'} \right. \nonumber\\
		&\quad\quad \left. \;\;+ \left( \int \!{\rm d}^3{\bm x} {}_{+} \psi^{{\rm (out)}\dagger}_{i} {}_- \psi^{\rm (in)}_{i'} \!\right) \hat{b}^{{\rm (in)}\dagger}_{i'} \right], \\
	\hat{a}^{({\rm out};1)}_{i} 
		&\!=\!  \sum_{i'} \!\!\!\!\!\!\!\! \int \left[ \left( \!- {\rm i}\int {\rm d}^4x {}_{+} \bar{\psi}^{{\rm (out)}}_{i} \Slash{\mathcal A} {}_+ \psi^{{\rm (in)}}_{i'} \!\right) \hat{a}^{{\rm (in)}}_{i'} \right.\nonumber\\
		&\quad\quad \left. \;\;+ \left( \!- {\rm i}\int \!{\rm d}^4x {}_{+} \bar{\psi}^{{\rm (out)}}_{i} \Slash{\mathcal A} {}_-\psi^{{\rm (in)}}_{i'} \!\right) \hat{b}^{{\rm (in)}\dagger}_{i'} \right] \!, \!\!
\end{align}
\end{subequations}
and
\begin{subequations}
\label{eq-12}
\begin{align}
	\hat{b}^{({\rm out};0)\dagger}_{i} 
		&\!=\! \sum_{i'} \!\!\!\!\!\!\!\! \int \left[ \left( \int \!{\rm d}^3{\bm x} {}_{-} \psi^{{\rm (out)}\dagger}_{i} {}_+ \psi^{\rm (in)}_{i'} \!\right) \hat{a}^{\rm (in)}_{i'} \right. \nonumber\\
		&\quad\quad \left. \;\;+ \left( \int \!{\rm d}^3{\bm x} {}_{-} \psi^{{\rm (out)}\dagger}_{i} {}_- \psi^{\rm (in)}_{i'} \!\right) \hat{b}^{{\rm (in)}\dagger}_{i'} \right], \\
	\hat{b}^{({\rm out};1)\dagger}_{i} 
		&\!=\!  \sum_{i'} \!\!\!\!\!\!\!\! \int \left[ \left( \!- {\rm i}\int {\rm d}^4x {}_{-} \bar{\psi}^{{\rm (out)}}_{i} \Slash{\mathcal A} {}_+ \psi^{{\rm (in)}}_{i'} \!\right) \hat{a}^{{\rm (in)}}_{i'} \right.\nonumber\\
		&\quad\quad \left. \;\;+\! \left( \!- {\rm i} \!\int \!{\rm d}^4x {}_{-} \bar{\psi}^{{\rm (out)}}_{i} \Slash{\mathcal A} {}_-\psi^{{\rm (in)}}_{i'} \!\right) \hat{b}^{{\rm (in)}\dagger}_{i'} \right] \!.\!\!
\end{align}
\end{subequations}
Therefore, the annihilation operator at the infinite future, $\hat{a}^{({\rm out})}_{i}$ or $\hat{b}^{({\rm out})}_{i}$, becomes a mixture of those at the infinite past, $\hat{a}^{({\rm in})}_{i}$ and $\hat{b}^{({\rm in})}_{i}$.  Notice that the annihilation operators at the infinite future, $\hat{a}^{({\rm out})}_{i}$ and $\hat{b}^{({\rm out})}_{i}$, contain the creation operators at the infinite past, $\hat{a}^{({\rm in})\dagger}_{i}$ and $\hat{b}^{({\rm out})\dagger}_{i}$, respectively.  This means that $\hat{a}^{({\rm out})}_{i}$ and $\hat{b}^{({\rm out})}_{i}$ do not annihilate the vacuum state at the infinite past, i.e., the in-vacuum expectation value of the number operator at the infinite future becomes non-vanishing.

We compute the (anti-)particle number $N$ ($\bar{N}$) at the infinite future produced from the in-vacuum.  Using Eqs.~(\ref{eq-11}) and (\ref{eq-12}), we can evaluate $N$ and $\bar{N}$ including the first order non-trivial correction by the perturbative field ${\mathcal A}_\mu$ as
\begin{align}
	\overset{(-)}{N} =  \sum_{i} \!\!\!\!\!\!\!\! \int \; \overset{(-)}{n}_i, \label{eq-13}
\end{align}
where ${n}_i$ and $\bar{n}_i$ are the distributions of particles and anti-particles per mode $i$, respectively, and are given by
\begin{subequations}
\label{eq-14}
\begin{align}
	n_{i} 
		&\equiv \frac{ \braket{ {\rm vac;in} | \hat{a}^{{\rm (out)}\dagger}_{i} \hat{a}^{{\rm (out)}}_{i}   | {\rm vac;in} } }{\braket{ {\rm vac;in} | {\rm vac;in} } } \nonumber\\
		&= \sum_{i'} \!\!\!\!\!\!\!\! \int \left| \left( \int {\rm d}^3{\bm x} {}_{+} \psi^{{\rm (out)}\dagger}_{i} {}_- \psi^{\rm (in)}_{i'} \right) \right. \nonumber\\
		&\quad\quad\quad \left. -{\rm i}e \left( \int {\rm d}^4x {}_{+} \bar{\psi}^{{\rm (out)}}_{i} \Slash{\mathcal A} {}_-\psi^{{\rm (in)}}_{i'} \right)  \right|^2 , \\
	\bar{n}_{i} 
		&\equiv \frac{ \braket{ {\rm vac;in} | \hat{b}^{{\rm (out)}\dagger}_{i} \hat{b}^{{\rm (out)}}_{i}   | {\rm vac;in} } }{\braket{ {\rm vac;in} | {\rm vac;in} } } \nonumber\\
		&= \sum_{i'} \!\!\!\!\!\!\!\! \int \left| \left( \int {\rm d}^3{\bm x} {}_{-} \psi^{{\rm (out)}\dagger}_{i} {}_+ \psi^{\rm (in)}_{i'} \right) \right. \nonumber\\
		&\quad\quad\quad \left. -{\rm i}e \left( \int {\rm d}^4x {}_{-} \bar{\psi}^{{\rm (out)}}_{i} \Slash{\mathcal A} {}_+\psi^{{\rm (in)}}_{i'} \right)  \right|^2 , \label{eq-14b}
\end{align}
\end{subequations}
with $\ket{\rm vac;in}$ being the vacuum at the infinite past defined as a state such that
\begin{align}
	0	= \hat{a}_i^{\rm (in)} \ket{\rm vac;in} 
		= \hat{b}_i^{\rm (in)} \ket{\rm vac;in}
		\ {\rm for\ any}\ i.  
\end{align}
The first term in Eq.~(\ref{eq-14}) accounts for the Schwinger mechanism because it is independent of the perturbation ${\mathcal A}_{\mu}$ and is driven only by the slow strong electromagnetic field $\bar{A}_{\mu}$.  The second term in Eq.~(\ref{eq-14}) accounts for one-photon pair production by the perturbation ${\mathcal A}_{\mu}$.  Note that the second term is distinct from the ordinary one-photon pair production since ours is non-perturbatively dressed by the slow strong electromagnetic field $\bar{A}_{\mu}$.  As shown below, this dressing enables us to describe the smooth interplay between the Schwinger mechanism and one-photon pair production as well as the dynamical assistance between the two mechanisms.

\subsection{Analytical discussion for a parallel strong electromagnetic field with a perturbation} \label{sec2b}

To get a qualitative understanding of the particle production, let us consider a spatially homogeneous system and a constant parallel strong electromagnetic field $\bar{\bm E} \parallel \bar{\bm B}$ superimposed by a fast weak perturbation ${\bm {\mathcal E}}$ pointing in the same direction.  Namely, we consider an electromagnetic field configuration,
\begin{align}
	{\bm E} = (0,0,\bar{E}+{\mathcal E}(x^0)),\ {\bm B}=(0,0,\bar{B}),  \label{eq_17}
\end{align}
which is realized by a gauge potential $A_{\mu} = \bar{A}_\mu + {\mathcal A}_{\mu}$ such that 
\begin{align}
	\left\{ \begin{array}{l}
		\displaystyle \bar{A}^\mu  		= (0,-\bar{B}x^2,0,-\bar{E}x^0) \vspace*{2mm}\\
		\displaystyle {\mathcal A}^{\mu}	= (0,0,0,-\int^{x^0}{\rm d}{x'}^0\;{\mathcal E}({x'}^0))
	\end{array} \right. .  \label{eq-17}
\end{align}
For simplicity, we assume $e\bar{E}, e\bar{B} > 0$.  Note that the time-dependence of ${\mathcal E}$ is arbitrary at this stage.

\subsubsection{Evaluation of the production number formula (\ref{eq-13})}

We analytically evaluate the production number formula (\ref{eq-13}) for the field configuration (\ref{eq-17}). 

First, we solve the Dirac equation (\ref{eq-6}) in the presence of the parallel strong electromagnetic field $\bar{A}_{\mu}$ (\ref{eq-17}).  This is analytically doable.  We find that there are four good quantum numbers $i = p_1,p_3,n$, and $s$ to label the mode function ${}_{\pm} \psi_{i}^{\rm (as)} = {}_{\pm} \psi_{p_1,p_3,n,s}^{\rm (as)}$; namely, the canonical momenta $p_1$ and $p_3$ (i.e., the eigen-values of the translation operators $-{\rm i}\partial_1$ and $-{\rm i}\partial_3$, respectively), the Landau level $n=0,1,2,\cdots \in {\mathbb N}$ (i.e., the remaining transverse momentum $p_2$ is quantized by the magnetic field), and spin $s=\pm 1$.  We can write down the mode function ${}_{\pm} \psi_{p_1,p_3,n,s}^{\rm (as)}$ explicitly as
\begin{align}
	\begin{pmatrix}
		{}_{+}\psi_{p_1,p_3,n,s}^{\rm (as)}(x) \\
		{}_{-}\psi_{p_1,p_3,n,s}^{\rm (as)}(x) 
	\end{pmatrix}
	&= 
	\begin{pmatrix}
		A^{\rm (as)}_{p_1,p_3,n,s}(x^0)   & B^{\rm (as)}_{p_1,p_3,n,s}(x^0) \\
		B^{{\rm (as)}*}_{p_1,p_3,n,s}(x^0) & -A^{{\rm (as)}*}_{p_1,p_3,n,s}(x^0) 
	\end{pmatrix} \nonumber\\
	&\quad\times
	\begin{pmatrix}
		U_{p_1,n,s}(x^2) \\
		V_{p_1,n,s}(x^2) 
	\end{pmatrix}
	\frac{{\rm e}^{{\rm i}p_1 x^1}{\rm e}^{{\rm i}p_3 x^3}}{2\pi} .  \label{eq11}
\end{align}
Here, the scalar functions $A^{\rm (as)}_{p_1,p_3,n,s}$ and $B^{\rm (as)}_{p_1,p_3,n,s}$ are given by 
\begin{subequations}
\begin{align}
	&\left\{\begin{array}{l}
		\!\!A^{\rm (in)}_{p_1,p_3,n,s}(x^0) \!\equiv\!  {\rm e}^{-\frac{{\rm i}\pi}{8}} {\rm e}^{-\frac{\pi}{8}  \frac{m_{\perp}^2}{e\bar{E}}} \frac{m_{\perp}}{\sqrt{2e\bar{E}}} \\
			\quad\quad\quad\quad\quad\quad\  \times  D_{\frac{\rm i}{2} \frac{m_{\perp}^2}{e\bar{E}}-1} \!\!\left(-{\rm e}^{-\frac{{\rm i}\pi}{4}} \sqrt{\frac{2}{e\bar{E}}} (p_3+e\bar{E}x^0) \right) \\
		\!\!B^{\rm (in)}_{p_1,p_3,n,s}(x^0) \!\equiv\!  {\rm e}^{+\frac{{\rm i}\pi}{8}} {\rm e}^{-\frac{\pi}{8} \frac{m_{\perp}^2 }{e\bar{E}}} \\
			\quad\quad\quad\quad\quad\quad\  \times D_{\frac{{\rm i}}{2} \frac{m_{\perp}^2}{e\bar{E}}} \!\!\left(-{\rm e}^{-\frac{{\rm i}\pi}{4}} \sqrt{\frac{2}{e\bar{E}}} (p_3 + e\bar{E}x^0) \right)
	\end{array}\right. \!\!\!, \label{eq25} \\
	&\left\{\begin{array}{l}
		\!\!A^{\rm (out)}_{p_1,p_3,n,s}(x^0) 
			\!\equiv\!  {\rm e}^{-\frac{{\rm i}\pi}{8}} {\rm e}^{-\frac{\pi}{8} \frac{m_{\perp}^2 }{e\bar{E}}}  \\
			\quad\quad\quad\quad\quad\quad\  \times D_{-\frac{{\rm i}}{2} \frac{m_{\perp}^2}{e\bar{E}}} \!\!\left({\rm e}^{\frac{{\rm i}\pi}{4}} \sqrt{\frac{2}{e\bar{E}}} (p_3 +e\bar{E}x^0) \right) \\
		\!\!B^{\rm (out)}_{p_1,p_3,n,s}(x^0) \!\equiv\!  {\rm e}^{+\frac{i\pi}{8}} {\rm e}^{-\frac{\pi}{8} \frac{m_{\perp}^2}{e\bar{E}}}  \frac{m_{\perp}}{\sqrt{2e\bar{E}}}  \\
			\quad\quad\quad\quad\quad\quad\  \times D_{-\frac{{\rm i}}{2} \frac{m_{\perp}^2}{e\bar{E}}-1} \!\!\left({\rm e}^{\frac{{\rm i}\pi}{4}} \sqrt{\frac{2}{e\bar{E}}} (p_3+e\bar{E}x^0) \right)
	\end{array}\right. \!\!\!,  \label{eq26}
\end{align}
\end{subequations}
with $D_{\nu}(z)$ being the parabolic cylinder function and $m_\perp$ being the transverse mass, 
\begin{align}
	m_{\perp} \equiv \sqrt{m^2 + e\bar{B}(2n+1-s) } .
\end{align}
The spinors $U_{p_1,n,s},V_{p_1,n,s}$ are given by
\begin{subequations}
\begin{align}
	U_{p_1,n,s}(x^2)	&\equiv u_{p_1,n}(x^2) \times \Gamma_s, \\
	V_{p_1,n,s}(x^2)	&\equiv \biggl[ \gamma^0 \frac{m}{m_\perp} u_{p_1,n}(x^2) \nonumber\\
					&\quad\  + \gamma^1 \frac{\sqrt{e\bar{B}(2n+1-s)}}{m_\perp} u_{p_1,n-s}(x^2) \biggl] \times \Gamma_s,
\end{align}
\end{subequations}
where 
\begin{align}
	\!\!u_{p_1,n} (x^2) \!\equiv\! \sqrt{\!\frac{L}{2\pi}} \!\left(\! \frac{e\bar{B}}{\pi} \!\right)^{\!\!\frac{1}{4}} \!\!\! \frac{1}{\sqrt{n!}} D_n \!\left(\! \sqrt{\!\frac{2}{e\bar{B}}}(p_1+e\bar{B} x^2)  \!\right)\! \!\label{eq27}
\end{align}
with $L$ being the system size in the $x^2$-direction and $\Gamma_s$ being an eigenvector of $\gamma^0 \gamma^3$ and $\gamma^1 \gamma^2$ satisfying
\begin{align}
	\gamma^0\gamma^3 \Gamma_s = + \Gamma_s,\ 
	\gamma^1\gamma^2 \Gamma_s = -{\rm i}s \Gamma_s,\ 
	\delta_{s,s'} = \Gamma_s^{\dagger} \Gamma_{s'}. \label{eq18}
\end{align}
Note that we normalized $A_{p_1,p_3,n,s}, B_{p_1,p_3,n,s}$, and $u_{p_1,n}$ as 
\begin{subequations}
\begin{align}
	&1 = |A^{\rm (as)}_{p_1,p_3,n,s}|^2 + |B^{\rm (as)}_{p_1,p_3,n,s}|^2, \label{eq18a}\\
	&\frac{L}{2\pi} \delta_{n,n'} = \int {\rm d}y\; u^{\dagger}_{p_1,n'} u_{p_1,n}, \label{eq18b}
\end{align}
\end{subequations}
so that the normalization condition (\ref{eq-7}) for the mode function ${}_\pm \psi^{\rm (as)}_{p_1,p_3,n,s}$ is satisfied as
\begin{align}
	\delta_{i,i'} = \delta_{s,s'}\times \delta(p_1-p'_1) \delta(p_3-p'_3) \times \frac{L}{2\pi} \delta_{n,n'}.  
\end{align}
Note that $(L/2\pi) \delta_{n,n'}$ corresponds to $\delta(p_2 - p'_2)$ in systems without any magnetic field.  Also, note that $A^{\rm (in)}_{p_1,p_3,n,s} \neq A^{\rm (out)}_{p_1,p_3,n,s}$ and $B^{\rm (in)}_{p_1,p_3,n,s} \neq B^{\rm (out)}_{p_1,p_3,n,s}$ because of the existence of the strong electric field.  This is essentially because the Dirac equation becomes time-dependent if $e\bar{E} \neq 0$ and then the eigen-function of the time-translation operator at the infinite past cannot be the one at the infinite future.  Physically, this means that the strong electric field supplies energy to the system, which mixes up particle and anti-particle modes.

In terms of the four good quantum numbers $i = p_1,p_3,n$, and $s$, we can express the mode integral (\ref{eq-5}) as
\begin{align}
	\hat{\psi}
		&=\! \sum_{s=\pm 1} \frac{2\pi}{L} \sum_{n=0}^{\infty} \int {\rm d}p_1 {\rm d}p_3\nonumber\\
		&\quad\!\! \times \!\!\left[ {}_{+}\!\psi_{p_1,p_3,n,s}^{\rm (as)} \hat{a}^{\rm (as)}_{p_1,p_3,n,s} \!+\! {}_{-}\!\psi_{p_1,p_3,n,s}^{\rm (as)} \hat{b}^{({\rm as})\dagger}_{-\!p_1,-\!p_3,n,-\!s}  \right] \!. \!\!\! \label{eqq4}
\end{align}
We assigned the minus signs in the labels of $\hat{b}^{({\rm as})}_{-p_1,-p_3,n,-s}$.  Those minus signs are necessary so that the creation operator $\hat{b}^{({\rm as})\dagger}_{p_1,p_3,n,s}$ correctly creates a one-particle state with quantum numbers $p_1,p_3,n$, and $s$.  One can check this statement by explicitly computing an expectation value of the corresponding operator at the asymptotic states $\hat{O}^{\rm (as)}$ [e.g., the canonical momentum operator $\hat{\psi}^{{\rm (as)}\dagger}(-{\rm i}{\bm \partial})\hat{\psi}^{\rm (as)}$] with respect to the one-particle state $\hat{b}^{({\rm as})\dagger}_{p_1,p_3,n,s} \ket{{\rm vac;as}}$ \cite{tan09}.  According to this assignment, we also have minus signs in the labels of $\bar{n}$ (\ref{eq-14b}) as
\begin{align}
	&\bar{n}_{-p_1,-p_3,n,-s} \nonumber\\
		&= \frac{ \braket{ {\rm vac;in} | \hat{b}^{{\rm (out)}\dagger}_{-p_1,-p_3,n,-s} \hat{b}^{{\rm (out)}}_{-p_1,-p_3,n,-s}   | {\rm vac;in} } }{\braket{ {\rm vac;in} | {\rm vac;in} } } \nonumber\\
		&= \sum_{s'=\pm 1} \frac{2\pi}{L} \sum_{n'=0}^{\infty} \int {\rm d}p'_1 {\rm d}p'_3\nonumber\\
		&\quad \times \left| \left( \int d^3{\bm x} {}_{-} \psi^{{\rm (out)}\dagger}_{p_1,p_3,n,s} {}_+ \psi^{\rm (in)}_{p'_1,p'_3,n',s'} \right) \right. \nonumber\\
		&\quad\quad\quad \left. -{\rm i}e \left( \int {\rm d}^4x {}_{-} \bar{\psi}^{{\rm (out)}}_{p_1,p_3,n,s} \Slash{\mathcal A} {}_+\psi^{{\rm (in)}}_{p'_1,p'_3,n',s'} \right)  \right|^2 .
\end{align}

Second, we evaluate the production number formula (\ref{eq-13}) using the analytical expression for the mode function ${}_\pm \psi^{\rm (as)}_{p_1,p_3,n,s}$ (\ref{eqq4}).  By substituting Eq.~(\ref{eqq4}) into Eq.~(\ref{eq-14}), one can show
\begin{align}
	&\!n_{p_1,p_3,n,s} 
	= \bar{n}_{-p_1,-p_3,n,-s} \nonumber\\
	&\!= \frac{V}{(2\pi)^3}  \exp\left[ -\pi \frac{ m_{\perp}^2}{e\bar{E}} \right] \nonumber\\
	&\quad\times\! \Biggl| 1 +  \frac{1}{2} \frac{  m_{\perp}^2 }{e\bar{E}} \!\int_0^{\infty} \!\!\!\!{\rm d}\omega \frac{\tilde{\mathcal E}(\omega)}{\bar{E}} \exp\!\left[ -\frac{\rm i}{4}\frac{\omega^2 + 4\omega p_3}{e\bar{E}}  \right] \nonumber\\
	&\quad\quad\quad\quad\quad\quad\quad\quad\quad \times\! {}_{1\!}\tilde{F}_1 \!\left(\! 1 \!-\! \frac{\rm i}{2} \frac{  m_{\perp}^2 }{e\bar{E}}; 2; \frac{\rm i}{2} \frac{\omega^2}{e\bar{E}} \right) \!\Biggl|^2\!\!, \!\! \label{eq61}
\end{align}
where $V$ is the space volume of the system, 
\begin{align}
	\tilde{\mathcal E}(\omega) \equiv \int {\rm d}x^0 {\rm e}^{-{\rm i}\omega x^0} {\mathcal E}(x^0) 
\end{align}
is the Fourier component of the perturbation ${\mathcal E}$, and ${}_1\tilde{F}_1(a;b;z)\equiv {}_1{F}_1(a;b;z)/\Gamma(b)$ is the regularized hypergeometric function.  Note that $n_{p_1,p_3,n,s} = \bar{n}_{-p_1,-p_3,n,-s}$ holds.  This is physically because a particle and an anti-particle are always produced as a pair from the vacuum and hence the total amounts of the charge, momentum, and spin of the pair should be vanishing.  By integrating Eq.~(\ref{eq61}) over the quantum numbers, we arrive at
\begin{align}
	\overset{(-)}{N}
	&= \sum_{s=\pm1} \frac{2\pi}{L} \sum_{n=0}^{\infty} \int {\rm d}p_1 {\rm d}p_3\; \overset{(-)}{n}_{p_1,p_3,n,s} \\
	&= m^4VT \times \frac{1}{4\pi^2} \frac{e\bar{E}}{m^2} \frac{e\bar{B}}{m^2} \sum_{s=\pm1} \sum_{n=0}^{\infty} \exp\left[ -\pi \frac{ m_\perp^2}{e\bar{E}} \right] \nonumber\\
	&\quad\times \Biggl[ 1 + \frac{2\pi}{T} \Biggl\{ \frac{1}{2} \frac{ m_{\perp}^2}{e\bar{E}}  \frac{\tilde{\mathcal E}(0)}{\bar{E}} + \frac{1}{4} \left( \frac{ m_{\perp}^2}{e\bar{E}} \right)^2 \int_0^{\infty} {\rm d}\omega \nonumber\\
	&\quad\quad\  \times \left| \frac{\tilde{\mathcal E}(\omega)}{\bar{E}} {}_1\tilde{F}_1 \left( 1- \frac{\rm i}{2} \frac{m_{\perp}^2}{e\bar{E}}; 2; \frac{\rm i}{2} \frac{\omega^2}{e\bar{E}} \right) \right|^2 \Biggl\} \Biggl] , \label{eq64}
\end{align}
where $T$ is the whole time interval, and we used $\int {\rm d}p_1 = e\bar{B} L$, which accounts for the degeneracy of the Landau level.  Notice that the total production number $\overset{(-)}{N}$ depends on the magnetic field $e\bar{B}$ through (i) the overall factor $e\bar{B}$, which comes from the phase-space under the Landau quantization $\int {\rm d}^2 {\bm p}_{\perp} \to (e\bar{B}/2\pi) \sum_n$; and (ii) the transverse mass $m_\perp$, which is discretized by the Landau quantization as $\sqrt{m^2+{\bm p}_{\perp}^2} \to \sqrt{m^2+e\bar{B}(2n+1-s)}=m_\perp$.  This means that the effect of the parallel strong magnetic field $e\bar{B}$ is just to discretize the transverse momentum via the Landau quantization, and one may say that the particle production (or the dynamically assisted Schwinger mechanism) is essentially unaffected by the presence of a strong magnetic field.  This is a reasonable result since magnetic fields cannot do work on a charged particle (i.e., cannot supply energy to vacuum fluctuations to produce real particles), and they are just able to change motion of the particle (i.e., can change momentum).  Nevertheless, the change of momentum can result in non-trivial physics consequences such as chirality production, which we discuss in Sec.~\ref{sec3}.

Below, let us consider some limiting situations to analytically understand basic features of the production number formula (\ref{eq64}).  In particular, we clarify the relationship between the formula (\ref{eq64}) and the well-established production formulas for the Schwinger mechanism and one-photon pair production in terms of the size of the physical parameters such as the electromagnetic field strength and the typical slowness/fastness of a perturbation.

\subsubsection{Dependence on the electric field $e\bar{E}$} \label{sec2b2}

The electric field strength $e\bar{E}$ controls the interplay between the Schwinger mechanism and one-photon pair production.

When the strong electric field becomes very strong $e\bar{E} \to \infty$, the Schwinger mechanism becomes free from the exponential suppression by mass and can produce particles abundantly.  As a result, the particle production is dominated by the Schwinger mechanism, i.e., the first term in Eq.~(\ref{eq61}) dominates the production as
\begin{align}
	n_{p_1,p_3,n,s} \!=\! \bar{n}_{-\!p_1,-\!p_3,n,-\!s} 
	\!\xrightarrow{\!\!e\bar{E}\to \infty\!\!}{}\!	
	\frac{V}{(2\pi)^{3\!}}  \exp\!\left[ -\pi \frac{ m_{\perp}^2}{e\bar{E}} \right] \!,\!
\end{align}
which yields
\begin{align}
	N=\bar{N}
	\xrightarrow{e\bar{E}\to \infty}{}&\;
	m^4VT \times \frac{1}{4\pi^2} \frac{e\bar{B}}{m^2} \frac{e\bar{E}}{m^2} \nonumber\\
	&\times \exp\left[ -\pi \frac{ m^2}{e\bar{E}} \right] \coth\left[ \pi \frac{e\bar{B}}{e\bar{E}}  \right] \nonumber\\
	\equiv&\; N_{\rm Schwinger}.  \label{e--q71} 
\end{align}
This agrees exactly with the Schwinger formula in the presence of a strong parallel magnetic field (\ref{eq2}) \cite{sch51, nik70, mar72, gav96, tan09, hid11a, hid11b, hat20}.

In case that the strong electric field is not so strong or absent $e\bar{E} \to 0$, the Schwinger mechanism does not take place.  The particle production is, then, driven solely by one-photon pair production by the perturbation $e{\mathcal E}$ without any modification from the strong electric field $e\bar{E}$ but with the Landau quantization by the strong magnetic field $e\bar{B}$.  Namely, only the second term in Eq.~(\ref{eq61}) contributes to the production and the distribution $\overset{(-)}{n}_{p_1,p_3,n,s}$ reads
\begin{align}
	&n_{p_1,p_3,n,s} = \bar{n}_{-p_1,-p_3,n,-s} \nonumber\\
	&\!\xrightarrow{e\bar{E}\to0}{} \frac{V}{(2\pi)^3} \frac{1}{4}\frac{m_{\perp}^2}{m_{\perp}^2+p_3^2} \frac{\left|e\tilde{\mathcal E}(2\sqrt{m_{\perp}^2+p_3^2})\right|^2}{m_{\perp}^2+p_3^2}.  \label{eq76}
\end{align}
By integrating Eq.~(\ref{eq76}), we find that the total production number $\overset{(-)}{N}$ is given by
\begin{align}
	N=\bar{N}
	\xrightarrow{e\bar{E}\to0}{} 
	&m^3V \times \frac{1}{16\pi^2} \frac{g\bar{B}}{m^2} \sum_{s=\pm1} \sum_{n=0}^{\infty} \int \frac{{\rm d}p_3}{m} \nonumber\\
	&\times \frac{m_{\perp}^2}{m_{\perp}^2+p_3^2} \frac{\left|e\tilde{\mathcal E}(2\sqrt{m_{\perp}^2+p_3^2})\right|^2}{m_{\perp}^2+p_3^2}.  \label{eq81}
\end{align}
Notice that the argument of $e\tilde{\mathcal E}$ is $2\sqrt{m_{\perp}^2+p_3^2}$, i.e., the perturbative electric field $e{\mathcal E}$ needs to supply energy $2 \times \sqrt{m_{\perp}^2+p_3^2}$ to produce a pair of particles\footnote{$n$-photon pair production processes ($n>1$) have lower thresholds $2 \times \sqrt{m_{\perp}^2+p_3^2} / n$.  Such higher order processes are parametrically suppressed by $(e{\mathcal E}/m^2)^{2n}$, so that they are negligible as long as the perturbation $e{\mathcal E}$ is sufficiently weak, and thus the dominant contribution to the production number always comes from the one-photon pair production ($n=1$).  }.  This means that one-photon pair production never occurs if the frequency $\Omega$ of the perturbative electric field is below the lowest energy level $\Omega < 2 m$, and that the production number increases sharply whenever the frequency matches the $n$-th energy level $\Omega =  2 m_\perp = 2\sqrt{m^2 + e\bar{B} (2n+1-s)}$.  In the presence of the strong electric field $e\bar{E}\neq 0$, these threshold behaviors are smeared because of the energy supply by $e\bar{E}$, which can be interpreted as the dynamical assistance to one-photon pair production by the Schwinger mechanism and vice versa, as we shall demonstrate explicitly in Sec.~\ref{sec2c}.

\subsubsection{Dependence on the magnetic field $e\bar{B}$} \label{sec2b3}

The effect of the magnetic field is just to discretize the energy level, and the energy difference among the levels increases with the magnetic field strength $e\bar{B}$.  Therefore, the number of the Landau levels that contribute to the particle production changes with $e\bar{B}$.  Unlike the electric field strength $e\bar{E}$, the magnetic field strength $e\bar{B}$ cannot control the interplay between the Schwinger mechanism and one-photon pair production because the relative size between the two mechanisms is independent of $e\bar{B}$.

For a very strong magnetic field $e\bar{B} \to \infty$, the lowest Landau level $n=0,s=\pm 1$ ($+$ for particle, and $-$ for anti-particle) dominates the production since the contributions from the higher Landau levels are exponentially suppressed by $e\bar{B}$.  Therefore, we have 
\begin{align}
	&\!\!\!n_{p_1,p_3,n,s} = \bar{n}_{-p_1,-p_3,n,-s} \nonumber\\
	&\!\!\!\!\!\xrightarrow{e\bar{B} \to \infty}{}
	\delta_{n,0} \delta_{s,+1} \times \frac{V}{(2\pi)^3}  \exp\left[ -\pi \frac{ m^2}{e\bar{E}} \right] \nonumber\\
	&\!\!\!\quad\quad\quad\times\! \Biggl| 1 +  \frac{1}{2} \frac{  m^2 }{e\bar{E}} \!\int_0^{\infty} \!\!\!\!{\rm d}\omega \frac{\tilde{\mathcal E}(\omega)}{\bar{E}} \exp\!\left[ -\frac{\rm i}{4}\frac{\omega^2 + 4\omega p_3}{e\bar{E}}  \right] \nonumber\\
	&\!\!\!\quad\quad\quad\quad\quad\quad\quad\quad\quad\quad\quad \times\! {}_{1\!}\tilde{F}_1 \!\left(\! 1 \!-\! \frac{\rm i}{2} \frac{  m^2 }{e\bar{E}}; 2; \frac{\rm i}{2} \frac{\omega^2}{e\bar{E}} \right) \!\!\Biggl|^2\!\!, \!\!
\end{align}
which yields
\begin{align}
	\overset{(-)}{N}
	&\!\!\xrightarrow{e\bar{B} \to \infty}{} \overset{(-)}{N}_{\rm LLL} \nonumber\\
	&\quad\quad\;\equiv m^4 VT \times \frac{1}{4\pi^2} \frac{e\bar{E}}{m^2} \frac{e\bar{B}}{m^2} \exp\left[ -\pi \frac{ m^2}{e\bar{E}} \right] \nonumber\\
	&\quad\quad\quad\;\times\! \Biggl[ 1 + \frac{2\pi}{T} \Biggl\{ \frac{1}{2} \frac{ m^2}{e\bar{E}}  \frac{\tilde{\mathcal E}(0)}{\bar{E}} + \frac{1}{4} \left( \frac{ m^2}{e\bar{E}} \right)^2 \int_0^{\infty}\!\! {\rm d}\omega \nonumber\\
	&\quad\quad\quad\quad\  \times\! \left| \frac{\tilde{\mathcal E}(\omega)}{\bar{E}} {}_1\tilde{F}_1 \left( \!1\!-\! \frac{\rm i}{2} \frac{m^2}{e\bar{E}}; 2; \frac{\rm i}{2} \frac{\omega^2}{e\bar{E}}\! \right)\! \right|^2 \!\Biggl\} \Biggl] .  \label{eq-34}
\end{align}
The strong magnetic field $e\bar{B}$ appears just as an overall factor in Eq.~(\ref{eq-34}) because of the enhancement of the phase-space $\propto e\bar{B}$.  Thus, the production number $\overset{(-)}{N}$ linearly increases with $e\bar{B}$ if the magnetic field is very strong.

The lowest Landau level production $\overset{(-)}{N}_{\rm LLL}$ is dominated by the Schwinger mechanism [i.e., first term in Eq.~(\ref{eq-34})] in the chiral limit $m \to 0$ as
\begin{align}
	\overset{(-)}{N}_{\rm LLL} 
	\xrightarrow{m\to 0}{}
	&VT \times \frac{e\bar{E}e\bar{B}}{4\pi^2}. \label{eq-35}
\end{align}
This is because the lowest Landau level is gapless in the chiral limit.  Therefore, the phase-space of particles is already occupied by the production from the Schwinger mechanism, and the additional production by the perturbative electric field is forbidden by the Pauli principle.

If the magnetic field is not so strong or absent $e\bar{B} \to 0$, the transverse momentum becomes continuous, and hence the summation over the Landau level $n$ may be written as an integration over the transverse momentum.  Namely, one may replace the summation $e\bar{B} \sum_{n}$ in Eq.~(\ref{eq64}) with an integration $\int {\rm d}p_\perp p_\perp$ with $p_\perp^2 \equiv e\bar{B}(2n+1-s)$ as
\begin{align}
	\overset{(-)}{N}
	&\!\!\xrightarrow{\!e\bar{B} \to 0\!}{} \!\!\sum_{s=\pm 1} \int {\rm d}^3{\bm p}\; \overset{(-)}{n}_{p_1,p_2,p_3,s} \nonumber\\
	&\!\!\!\!\quad\quad =\! m^4 VT \!\times\! \frac{1}{2\pi^2} \frac{e\bar{E}}{m^2} \int_0^{\infty} \frac{{\rm d}p_{\perp}\;p_\perp}{m^2} \!\exp\!\left[ -\pi \frac{ m^2 \!+\! p_\perp^2}{e\bar{E}} \right] \nonumber\\
	&\quad\quad\times\! \Biggl[ 1 \!+\! \frac{2\pi}{T} \!\Biggl\{ \!\frac{1}{2} \!\frac{ m^2\!+\!p_{\perp}^2}{e\bar{E}} \! \frac{\tilde{\mathcal E}(0)}{\bar{E}} \!+\! \frac{1}{4}\! \left(\! \frac{ m^2\!+\!p_{\perp}^2}{e\bar{E}} \!\right)^{\!2} \!\!\!\int_0^{\infty} \!\!\!\!{\rm d}\omega \nonumber\\
	&\quad\quad\quad\  \times \! \left| \frac{\tilde{\mathcal E}(\omega)}{\bar{E}} {}_1\tilde{F}_1 \!\left(\! 1\!-\! \frac{\rm i}{2} \!\frac{m^2\!+\!p_{\perp}^2}{e\bar{E}}; \!2; \frac{\rm i}{2} \!\frac{\omega^2}{e\bar{E}} \!\right) \!\right|^2 \!\!\Biggl\} \Biggl] . \! \label{eq-36}
\end{align}
Equation~(\ref{eq-36}) exactly agrees with the result of Ref.~\cite{fk}, in which the production number formula (\ref{eq64}) for a purely slow strong electric field was derived.

Before continuing, it is instructive for later discussions to take the $e\bar{E} \to 0$ limit of Eq.~(\ref{eq-36}), i.e., to derive a formula for one-photon pair production without any strong electromagnetic field.  We find
\begin{align}
	\!\!\!N=\bar{N}
	\xrightarrow{e\bar{E}, e\bar{B} \to 0}{} 
	&m^3V \times \frac{1}{4\pi^2} \int_0^{\infty} \frac{{\rm d}p}{m} \frac{p^2}{m^2}  \nonumber\\
	&\times \frac{m^2+\frac{2}{3}p^2}{m^2+p^2} \frac{\left|e\tilde{\mathcal E}(2\sqrt{m^2+p^2})\right|^2}{m^2+p^2} \nonumber\\
	\equiv&\; N_{\rm one-photon}, \label{eq-39}
\end{align}
where $p^2 \equiv p_\perp^2 + p_3^2$.  In contrast to the one-photon pair production in the presence of $e\bar{B} \neq 0$ [Eq.~(\ref{eq81})], the energy level is continuous in Eq.~(\ref{eq-39}).  Hence, the production number does not exhibit any threshold behaviors in terms of the frequency of the perturbation.  In other words, addition of a strong parallel magnetic field results in sharp peak structures in the production number as a function of the frequency because of the Landau quantization.

\subsubsection{Dependence on the frequency of the perturbation $\Omega$} \label{sec2b4}

The typical frequency of a perturbation, which we write as $\Omega$, can control the interplay between the Schwinger mechanism and one-photon pair production.  Intuitively, the energy supplied by a perturbation via a one-photon process is $\Omega$.  Therefore if the supplied energy by a perturbation $\Omega$ becomes much larger (smaller) than that by the strong electric field $e\bar{E}$, one-photon pair production (the Schwinger mechanism) should dominate the production.  The strong magnetic field is not essential for this interplay since it cannot supply energy as discussed in the previous subsections.

Suppose that the frequency is very small $\Omega \to 0$.  This is equivalent to assuming
\begin{align}
	{\mathcal E}(x^0) = {\mathcal E}_0\ \Leftrightarrow\ \tilde{\mathcal E}(\omega) = {\mathcal E}_0 \times 2\pi \delta (\omega).  \label{eq_44}
\end{align}
Then, we can analytically carry out the $\omega$-integration in Eq.~(\ref{eq61}) as
\begin{align}
	&\!n_{p_1,p_3,n,s} 
	= \bar{n}_{-p_1,-p_3,n,-s} \nonumber\\
	&\!\!\xrightarrow{\Omega \to 0}{} \frac{V}{(2\pi)^3}  \exp\left[ -\pi \frac{ m_{\perp}^2}{e\bar{E}} \right]\Biggl| 1 +  \frac{\pi}{2} \frac{  m_{\perp}^2 }{e\bar{E}} \frac{{\mathcal E}_0}{\bar{E}} \Biggl|^2 \nonumber\\
	&\quad\;  =  \frac{V}{(2\pi)^3}  \exp\left[ -\pi \frac{ m_{\perp}^2}{e(\bar{E} + {\mathcal E}_0 ) } \right]^2 + {\mathcal O}(({\mathcal E}_0/\bar{E})^2).  
\end{align}
Therefore, we can evaluate the total production number $\overset{(-)}{N}$ as
\begin{align}
	\!\!\!N\!=\!\bar{N}
	\!\xrightarrow{\!\Omega \to 0\!}{}&
	m^4VT \!\times\! \frac{1}{4\pi^2}\!\frac{e\bar{B}}{m^2} \!\frac{e(\bar{E}\!+\!{\mathcal E}_0)}{m^2} \nonumber\\
	&\!\times\! \exp\!\left[ -\pi \frac{ m^2}{e(\bar{E}\!+\!{\mathcal E}_0)} \right] \coth\!\left[ \pi \frac{e\bar{B}}{e(\bar{E}\!+\!{\mathcal E}_0)}  \right]  \label{eq-44}
\end{align}
where ${\mathcal O}(({\mathcal E}_0/\bar{E})^2)$-correction is neglected.  By noting that the total electric field strength for $\Omega \to 0$ (\ref{eq_44}) is $E = \bar{E}+{\mathcal E}_0$, we see that Eq.~(\ref{eq-44}) agrees with the well-established Schwinger formula (\ref{e--q71}).  Therefore, the particle production is, indeed, dominated by the Schwinger mechanism if the frequency $\Omega$ is small.

If the frequency is very large $\Omega \to \infty$, the $\omega$-integration in Eq.~(\ref{eq61}) is dominated by $\omega = \Omega \to \infty$ modes.  Thus, we find
\begin{align}
	&n_{p_1,p_3,n,s} 
	= \bar{n}_{-p_1,-p_3,n,-s} \nonumber\\
	&\!\!\xrightarrow{\Omega \to \infty}{} \frac{V}{(2\pi)^3} \Biggl| \exp\left[ - \frac{\pi}{2} \frac{ m_{\perp}^2}{e\bar{E}} \right] \nonumber\\
	&\quad\quad\quad\quad\quad\quad + \frac{1}{2} \frac{m_\perp}{\sqrt{m_\perp^2+p_3^2}} \frac{e\tilde{\mathcal E}(2\sqrt{m_\perp^2+p_3^2})}{\sqrt{m_\perp^2+p_3^2}  } \Biggl|^2 \nonumber\\
	&\!\!\xrightarrow{e\bar{E} \lesssim m_\perp^2}{} \frac{V}{(2\pi)^3} \times \frac{1}{4} \frac{m_\perp^2}{m_\perp^2+p_3^2} \frac{\left| e\tilde{\mathcal E}(2\sqrt{m_\perp^2+p_3^2}) \right|^2}{m_\perp^2+p_3^2 } , \label{eq-45}
\end{align}
which yields
\begin{align}
	&N=\bar{N} \nonumber\\
	&\!\!\xrightarrow{\Omega \to \infty, e\bar{E} \lesssim m_\perp^2}{} 
	m^3V \times \frac{1}{16\pi^2} \frac{e\bar{B}}{m^2} \sum_{s=\pm1} \sum_{n=0}^{\infty} \int \frac{{\rm d}p_3}{m} \nonumber\\
	&\quad\quad\quad\quad\quad\quad \times \frac{m_{\perp}^2}{m_{\perp}^2+p_3^2} \frac{\left|e\tilde{\mathcal E}(2\sqrt{m_{\perp}^2+p_3^2})\right|^2}{m_{\perp}^2+p_3^2}.  \label{eq--46}
\end{align}
To get Eq.~(\ref{eq-45}), we neglected the first term in the first line (i.e., the contribution from the Schwinger mechanism), which is exponentially suppressed by $e\bar{E}/m_\perp^2$, assuming that $e\bar{E}$ is not so strong $e\bar{E} \lesssim m_\perp^2$.  If this condition is satisfied, Eq.~(\ref{eq--46}) agrees with Eq.~(\ref{eq81}), i.e., the production is dominated by one-photon pair production.  If $e\bar{E}$ is very strong $e\bar{E} \gtrsim m_\perp^2$, the strong electric field can supply very large energy to the vacuum, so that the Schwinger mechanism always dominates the production no matter how fast a perturbation is, as we discussed in Sec.~\ref{sec2b2}.

\subsection{Quantitative discussion for a parallel strong electromagnetic field with a monochromatic perturbation} \label{sec2c}

To understand more detailed features of the particle production, let us consider, as a demonstration, the parallel electromagnetic field configuration (\ref{eq-17}) with a monochromatic perturbation,
\begin{align}
	&{\mathcal E}(x^0) = {\mathcal E}_0 \cos(\Omega x^0 + \phi) \nonumber\\
	&\Leftrightarrow\ 
	\tilde{\mathcal E}(\omega) = \pi {\mathcal E}_0 \sum_{\pm} {\rm e}^{\pm {\rm i}\phi} \delta(\Omega \mp \omega) .   \label{eq-46}
\end{align}
For this perturbation, one can analytically carry out the $\omega$-integration in the production number formula (\ref{eq64}) as
\begin{align}
	\!\!\!N \!=\! \bar{N}
	&\!=\! m^4VT \!\times\! \frac{1}{4\pi^2} \!\frac{e\bar{E}}{m^2} \!\frac{e\bar{B}}{m^2} \!\sum_{s=\pm1} \!\sum_{n=0}^{\infty} \exp\!\left[ -\pi \frac{ m_\perp^2}{e\bar{E}} \right] \nonumber\\
	&\quad\!\times\! \Biggl[ 1 \!+\! \Biggl| \frac{\pi}{2} \frac{ m_{\perp}^2}{e\bar{E}}\frac{{\mathcal E}_0}{\bar{E}} {}_1\tilde{F}_1\! \left( 1\!-\! \frac{\rm i}{2}\! \frac{m_{\perp}^2}{e\bar{E}}; \!2; \frac{\rm i}{2} \!\frac{\Omega^2}{e\bar{E}} \right)\! \Biggl|^2 \Biggl] .\!\! \label{eq-47}
\end{align}
Note that the phase $\phi$ is unimportant in the total production number $\overset{(-)}{N}$ (\ref{eq-47}), but can affect the distribution $\overset{(-)}{n}$ through the quantum interference between the Schwinger mechanism and one-photon pair production \cite{dum10, dum11, heb09, gre17, fk, fk2}.  Below, we explicitly carry out the $n$- and $s$-summations in Eq.~(\ref{eq-47}) to quantitatively discuss how the interplay and dynamical assistance between the Schwinger mechanism and one-photon pair production occur.

\begin{figure*}[!t]
\begin{center}
\includegraphics[width=0.49\textwidth]{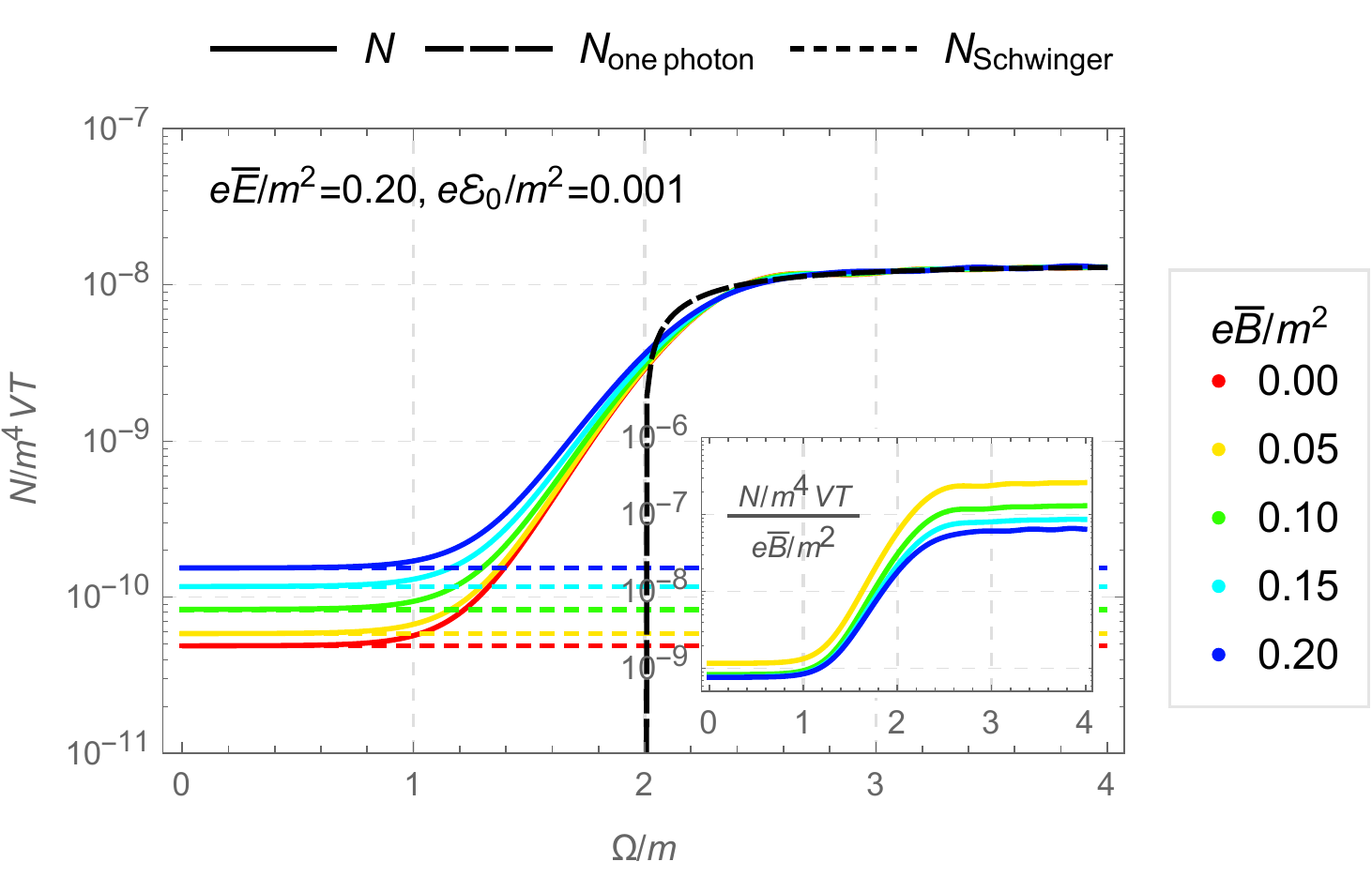}
\includegraphics[width=0.49\textwidth]{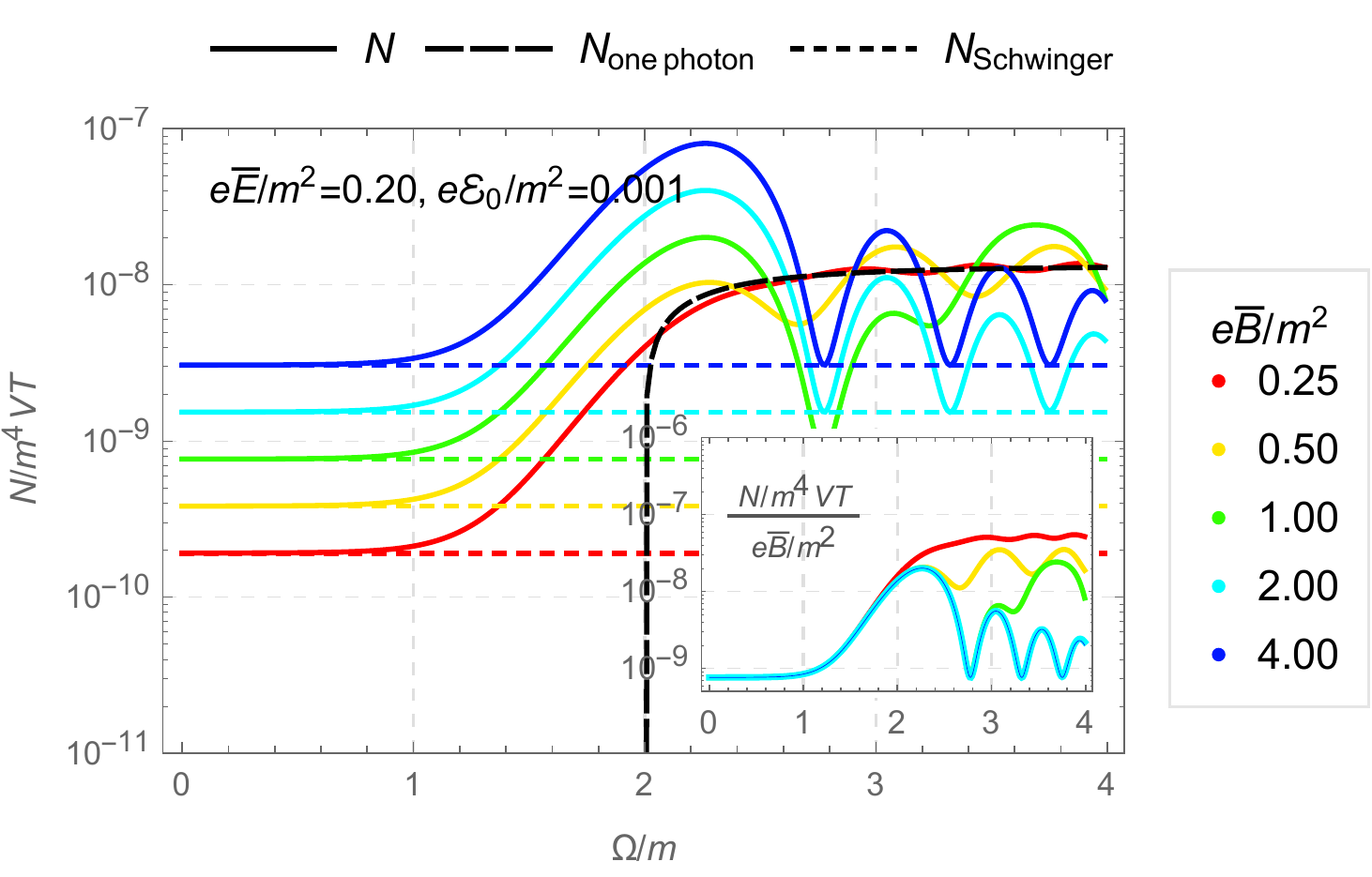}
\caption{\label{fig-2} The total production number $N$ (\ref{eq-47}) for the monochromatic perturbation (\ref{eq-46}) as a function of the frequency $\Omega/m$ for various values of the magnetic field strength $e\bar{B}/m^2 = 0.00, 0.05, 0.10, 0.15, 0.20$ (left) and $0.25, 0.50, 1.00, 2.00, 4.00$ (right).  The small panels show the total production number scaled by the magnetic field strength $N/(e\bar{B}/m^2)$.  The other parameters are fixed as $e\bar{E}/m^2=0.20$ and $e{\mathcal E}_0/m^2=0.001$.  For comparison, the one-photon pair production formula $N_{\rm one-photon}$ (\ref{eq-39}) and the Schwinger formula $N_{\rm Schwinger}$ (\ref{e--q71}) are plotted by the dashed and dotted lines, respectively.  }
\end{center}
\end{figure*}

Figure~\ref{fig-2} shows the total production number $N$ (\ref{eq-47}) as a function of the frequency of the perturbation $\Omega$ for various values of the magnetic field strength $e\bar{B}$.  We confirm that the particle production is dominated by the Schwinger mechanism for small frequency $\Omega \lesssim 2m$ and by one-photon pair production for large frequency $\Omega \gtrsim 2m$ as we discussed analytically in Sec.~\ref{sec2b4}.  At the intermediate frequency $\Omega \sim 2m$, the particle production becomes more abundant than what is expected from the Schwinger mechanism or one-photon pair production separately.  This is the dynamical assistance between the two production mechanisms.  Intuitively, the Schwinger mechanism is enhanced because the mass gap is reduced by the energy supply from the one-photon interaction and vice versa, as we explained in Fig.~\ref{fig-1}.

\begin{figure*}[!t]
\begin{center}
\includegraphics[width=0.49\textwidth]{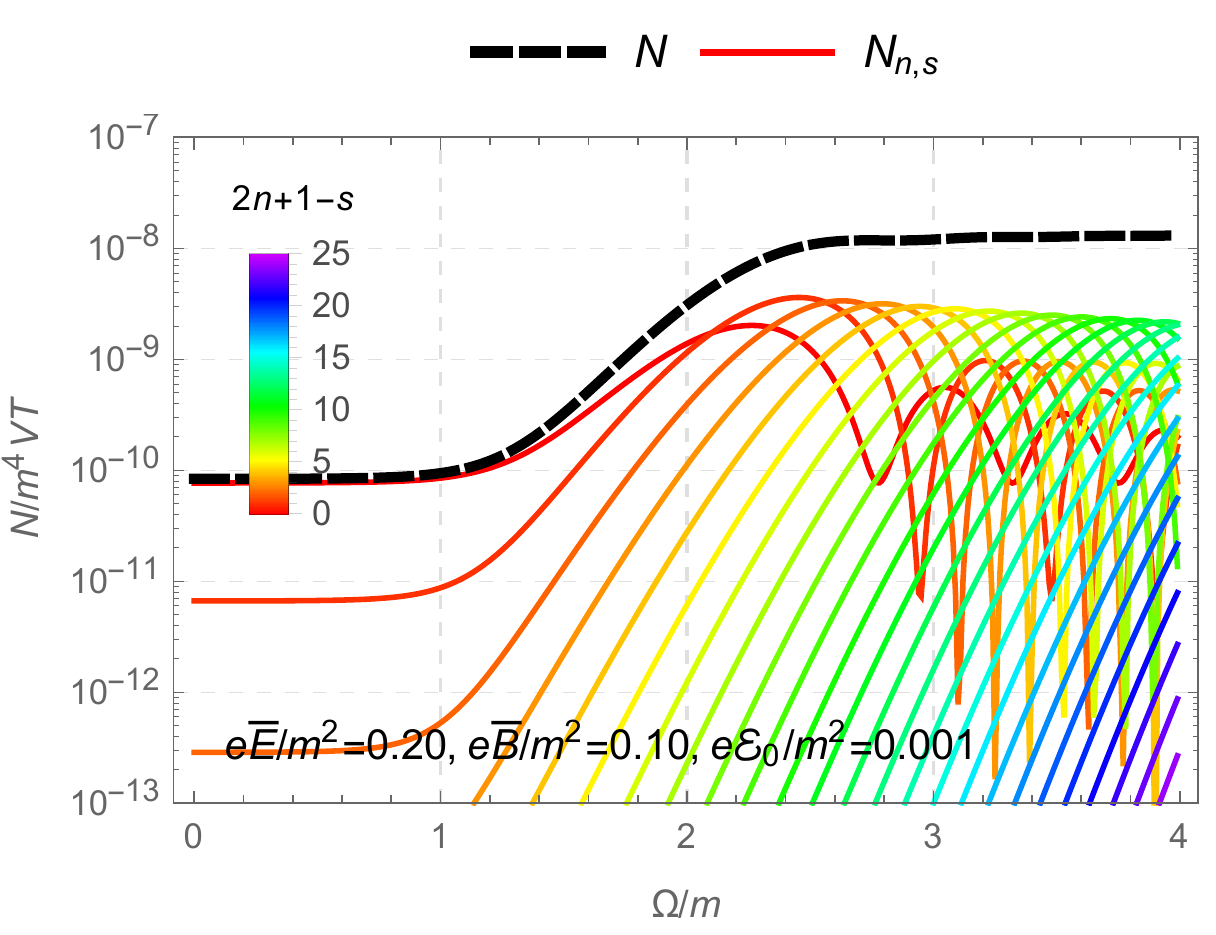}
\includegraphics[width=0.49\textwidth]{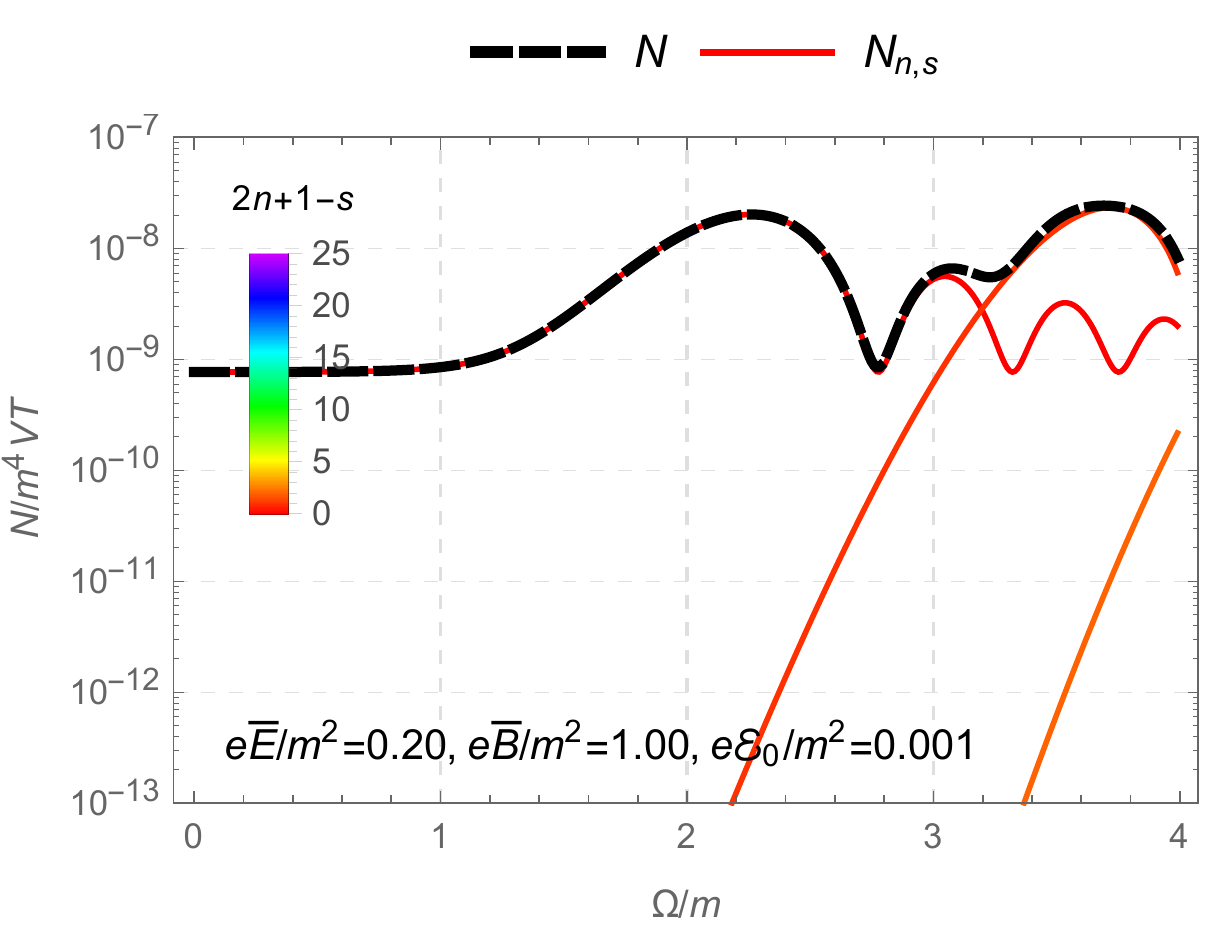}
\caption{\label{fig-3} The total production number $N = \sum_{n,s} N_{n,s}$ (\ref{eq-47}) (black dashed line) and the contribution from each Landau level $N_{n,s}$ (colored lines) for various values of $2n+1-s$ as a function of the frequency $\Omega/m$.  The magnetic field strength $e\bar{B}$ is different between the two panels as $e\bar{B}/m^2=0.10$ (left) and $1.00$ (right).  The other parameters are fixed as $e\bar{E}/m^2=0.20$ and $e{\mathcal E}_0/m^2=0.001$.  }
\end{center}
\end{figure*}

The frequency $\Omega$-dependence of one-photon pair production (i.e., the particle production for large frequency $\Omega \gtrsim 2m$) dramatically changes with increasing the magnetic field strength $e\bar{B}$.  Namely, the production number as a function of $\Omega$ becomes flat (oscillating) if $e\bar{B}$ is small (large).  This change occurs because the number of the Landau levels that contribute to the production changes depending on the size of $e\bar{B}$ (cf. Sec.~\ref{sec2b3}).  As shown in Fig.~\ref{fig-3}, the contribution from each Landau level $N_{n,s}$, 
\begin{align}
	N	\!=\! \sum_{s} \frac{2\pi}{L} \sum_{n} \!\int\! {\rm d}p_x \!\int\! {\rm d}p_z\; n_{p_x,p_z,n,s} 
		\!\equiv\! \sum_{n,s} N_{n,s},  \label{eq---49}
\end{align}
always has an oscillating dependence on $\Omega$.  This can be understood as an analog of the Franz-Keldysh oscillation in semi-conductor physics \cite{tah63, cal63, fk, fk2}, which occurs because of a quantum reflection in the presence of a tilted vacuum band structure under a strong electric field \cite{fk}.  The quantum reflection is a counter phenomenon of the quantum tunneling responsible for the Schwinger mechanism.  $N_{n,s}$ becomes significant above the energy threshold $\Omega \gtrsim 2 m_{\perp} = 2 \sqrt{m^2 + e\bar{B}(2n+1-s)}$.  Therefore, many Landau levels can contribute to the production for small $e\bar{B}$, which cancels the oscillation of each contribution.  On the other hand, such a cancellation does not take place for large $e\bar{B}$, for which only a small number of Landau levels can contribute to the production.

For very large magnetic field strength $e\bar{B} \gtrsim \Omega^2, m^2, e\bar{E}$, the lowest Landau level dominates the production $N \to N_{\rm LLL}$ [see Eq.~(\ref{eq-34})].  Then, the production number becomes just proportional to $e\bar{B}$ as shown in the small panels in Fig.~\ref{fig-2}.  The lowest Landau contribution $N_{\rm LLL}$ becomes the largest at around the threshold value $\Omega \sim 2 m $, which is essential for the dynamical enhancement of chirality production, as we discuss in Sec.~\ref{sec3}.

\begin{figure*}[!t]
\begin{center}
\includegraphics[width=0.49\textwidth]{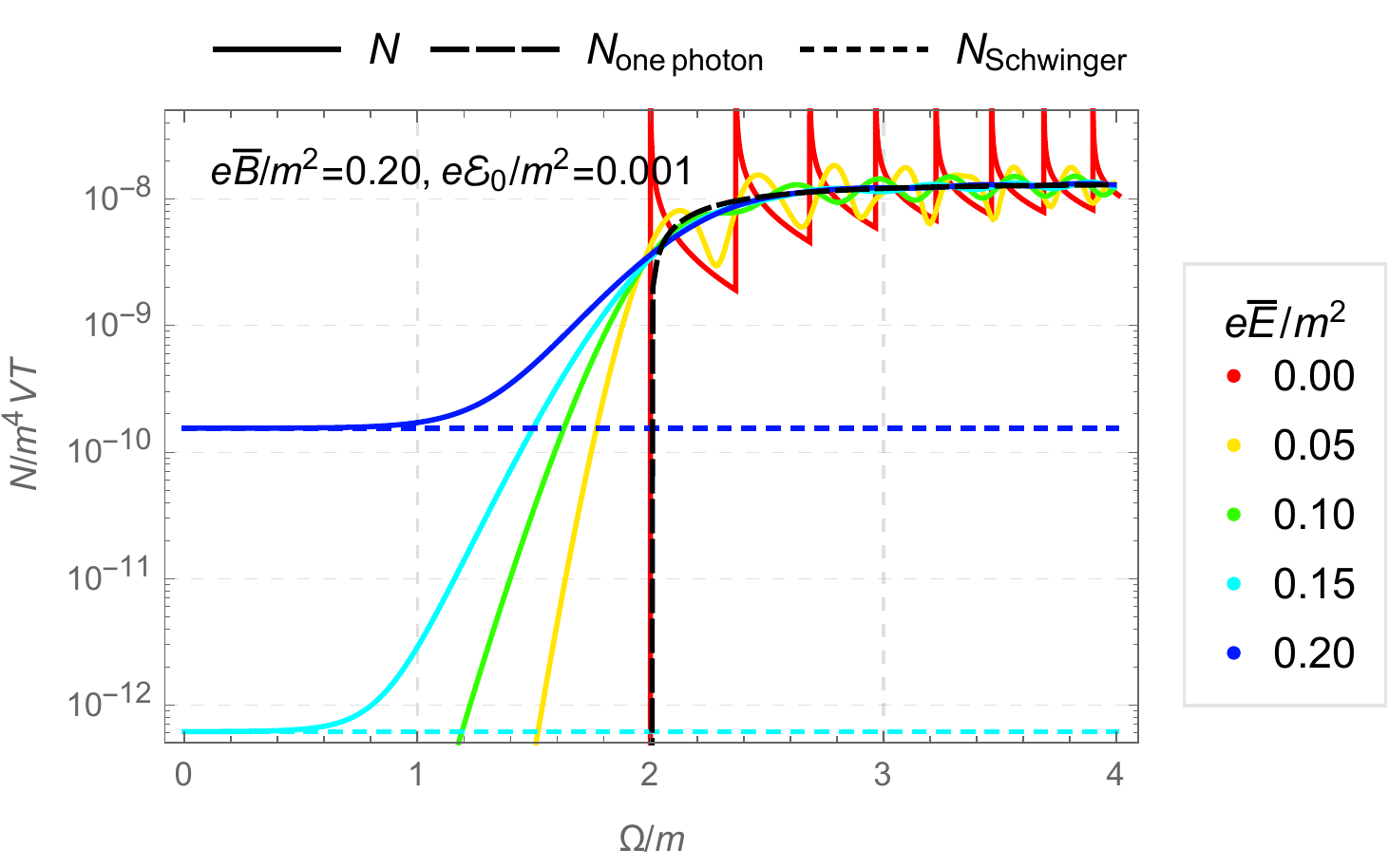}
\includegraphics[width=0.49\textwidth]{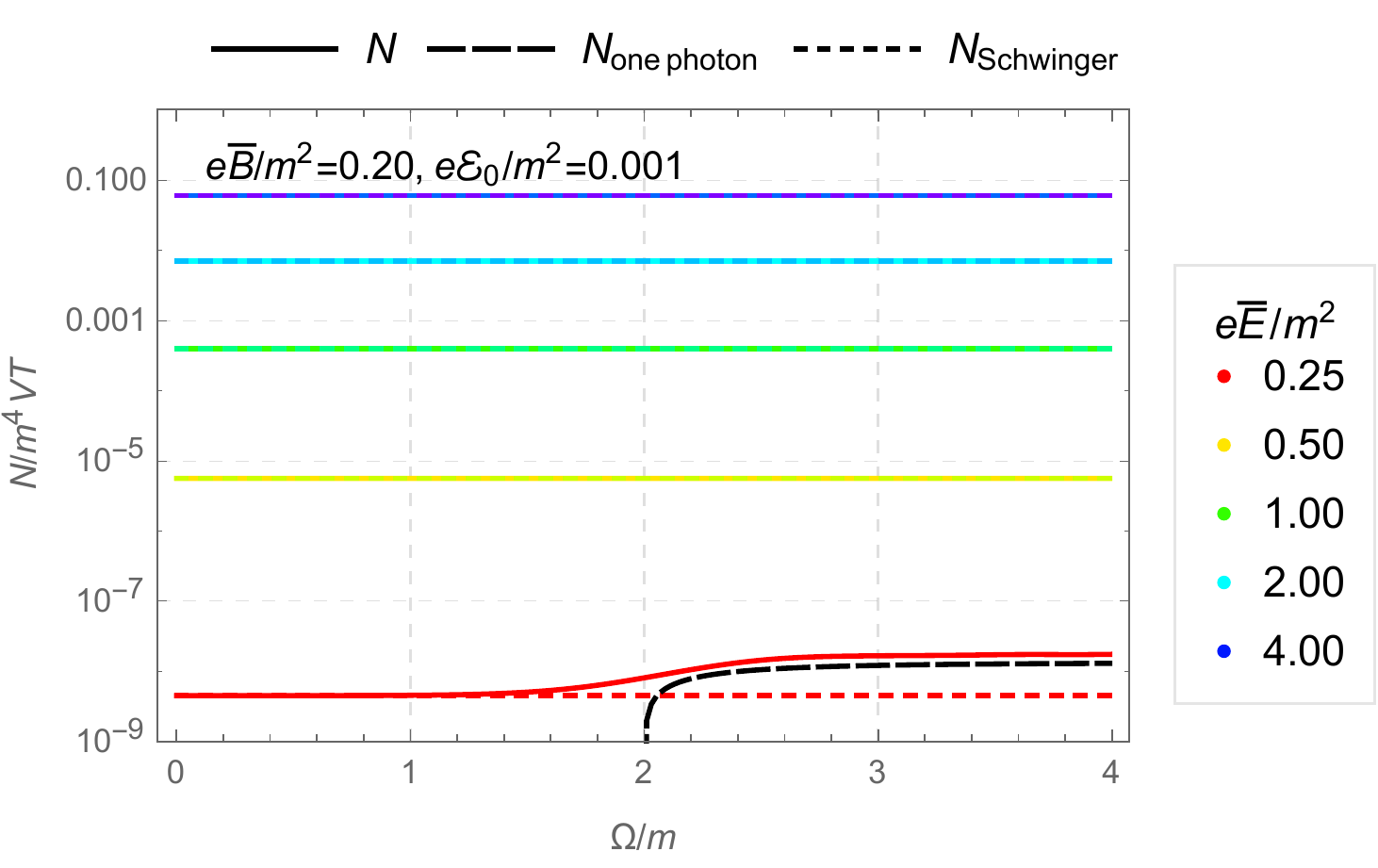}
\caption{\label{fig-4} The total production number $N$ (\ref{eq-47}) for the monochromatic perturbation (\ref{eq-46}) as a function of the frequency $\Omega/m$ for various values of the electric field strength $e\bar{E}/m^2 = 0.00, 0.05, 0.10, 0.15, 0.20$ (left) and $0.25, 0.50, 1.00, 2.00, 4.00$ (right).  The other parameters are fixed as $e\bar{B}/m^2=0.20$ and $e{\mathcal E}_0/m^2=0.001$.  For comparison, the one-photon pair production formula $N_{\rm one-photon}$ (\ref{eq-39}) and the Schwinger formula $N_{\rm Schwinger}$ (\ref{e--q71}) are plotted by the dashed and dotted lines, respectively.  Note that the vertical scale is different between left and right panels.  }
\end{center}
\end{figure*}

Figure~\ref{fig-4} shows the total production number $N$ (\ref{eq-47}) for various values of the electric field strength $e\bar{E}$.  The behavior of the particle production dramatically changes depending on the size of $e\bar{E}$ (cf. Sec.~\ref{sec2b2}).  For a weak electric field $e\bar{E} \lesssim m^2, e\bar{B}$ (the left panel of Fig.~\ref{fig-4}), the interplay between the Schwinger mechanism and one-photon pair production occurs and is controlled by the size of the frequency $\Omega$.  The particle production for large $\Omega$ shows up an oscillating behavior.  For vanishing electric field strength $e\bar{E}/m^2=0$, the particle production for large $\Omega$ is dominated by one-photon pair production in the presence of a strong magnetic field (\ref{eq81}), which is sharply peaked at the thresholds $\Omega = 2m_{\perp}$.  The threshold behavior is smeared by the electric field $e\bar{E} \neq 0$ since it supplies energy.  For large electric field strength $e\bar{E} \gtrsim m^2, e\bar{B}$ (the right panel of Fig.~\ref{fig-4}), the production is dominated by the Schwinger mechanism no matter how large the frequency $\Omega$ is, and hence the production number becomes constant in $\Omega$.  This is because the Schwinger mechanism becomes free from the exponential suppression and always surpasses one-photon pair production, which is suppressed by the power of $e\mathcal{E}/m^2$.  Note that the production number at $e\bar{E}/m^2=0.25$ (the red line in the right panel of Fig.~\ref{fig-4}) for large $\Omega$ does not coincide with the one-photon pair production formula (the dashed black line in the plot).  This is because even though the electric field strength $e\bar{E}/m^2=0.25$ is still weak for the Schwinger mechanism to dominate the production for any values of $\Omega$, it is enough strong that the Schwinger mechanism becomes comparable to one-photon pair production.  Thus, one cannot drop the contribution from the Schwinger mechanism as in the second line of Eq.~(\ref{eq-45}), and the production number is given by the sum of the two mechanisms as $N = N_{\rm one-photon} + N_{\rm Schwinger}$.

\section{Chirality production} \label{sec3}

In this section, we discuss how the dynamically assisted Schwinger mechanism affects chirality production.  Namely, we consider the slow strong parallel electromagnetic field superimposed by a fast weak perturbation (\ref{eq_17}) and derive an analytical formula for chirality production by explicitly evaluating an in-in vacuum expectation value of the chirality operator (see Sec.~\ref{sec3a}).  Then, we use that formula to show that the dynamical assistance can enhance chirality production by many orders of the magnitude, and that there exists an optimal frequency for a perturbation to maximize chirality production (see Sec.~\ref{sec3b}).

\subsection{Relation to the production number} \label{sec3a}

We analytically show that chirality production is related to the lowest Landau level production $N_{\rm LLL}$ and write down a formula for chiral charge produced from the vacuum in the presence of the slow strong parallel electromagnetic field superimposed by a perturbation (\ref{eq_17}) by using the production number formula derived in Sec.~\ref{sec2}.

The chiral charge at out-state $Q_5$ produced from an initial vacuum state $\ket{\rm vac;in}$ is defined as
\begin{align}
	\!Q_5 \!\equiv\!\!\!  \lim_{x^0 \!\to \infty} \!\! \int \!{\rm d}^3{\bm x} \braket{  {\rm vac;in} | \!: \hat{\bar{\psi}}^{\rm (out)}\gamma^0 \gamma_5 \hat{\psi}^{\rm (out)}   : \!| {\rm vac;in}} \!, \!\!\label{eqq169}
\end{align}
where $\gamma_5 \equiv {\rm i} \gamma^0 \gamma^1\gamma^2\gamma^3$ and $\braket{: \cdots :}$ is the normal ordering\footnote{Precisely speaking, normal ordering with respect to ``out-state'' operators.  } (e.g., $\braket{ : \hat{o}_1 \hat{o}_2^{\dagger} : } = - \braket{ \hat{o}_2^{\dagger} \hat{o}_1 }$), which is introduced so as to kill the ultraviolet divergence (i.e., subtract the unphysical vacuum contributions).  Note that $Q_5$ is defined as an {\it in-in} expectation value, whose significance and relation to the in-out expectation value were clarified within the proper-time formalism in Ref.~\cite{cop18}.

We can analytically evaluate Eq.~(\ref{eqq169}).  By substituting the mode expansion (\ref{eqq4}) into Eq.~(\ref{eqq169}), we obtain
\begin{align}
		\!\!Q_5
		&\!=\! \frac{(2\pi)^3}{V} \sum_{s} \frac{2\pi}{L} \sum_{n} \int {\rm d}p_x \int {\rm d}p_z\;n_{p_x,p_z,n,s}   \nonumber\\
			&\quad\!\!\!\times\!\! \lim_{t \to \infty}\! \sum_{\pm} (\pm 1) \!\left( \!\int \!\!{\rm d}^3{\bm x} {}_{\pm\!}\bar{\psi}^{\rm (out)}_{p_x,p_z,n,s\!} \gamma^0 \! \gamma_5 {}_{\pm\!} \psi^{\rm (out)}_{p_x,p_z,n,s\!} \!\right)\!, \!\!\! \label{eq3a2}
\end{align} 
where we used $n_{p_x,p_z,n,s} = \bar{n}_{-p_x,-p_z,n,-s}$ and dropped a severely oscillating factor $\lim_{x^0 \to \infty} {}_-\bar{\psi}^{\rm (out)}_{p_x,p_z,n,s} \gamma^0 \gamma_5 {}_+\psi^{\rm (out)}_{p_x,p_z,n,s} \propto \lim_{x^0 \to \infty} {\rm e}^{-2{\rm i}\sqrt{m_{\perp}^2+p_z^2}x^0} \to 0$ because of the ${\rm i}\epsilon$-prescription of quantum field theory.  Then, by using the analytical expression for the mode function ${}_\pm \psi^{\rm (as)}_{p_x,p_z,n,s}$ (\ref{eq11}), we can explicitly evaluate the matrix element in Eq.~(\ref{eq3a2}), and the result reads
\begin{align}
	Q_5	= \sum_{s}\sum_{n} N_{n,s} \times 2s \frac{m^2}{m_{\perp}^2} 
		= 2N_{\rm LLL}.  \label{eqq188}
\end{align}
To get the second equality, we used $N_{n,s=-1}=N_{n+1,s=+1}$.  The higher Landau level contributions cancel each other, and we are left only with the lowest Landau level contribution $2N_{\rm LLL} = N_{n=0,s=+1} + \bar{N}_{n=0,s=-1}$.  Equation~(\ref{eqq188}) suggests that chirality production should be enhanced if the lowest Landau level production $N_{\rm LLL}$ is enhanced by, e.g., the dynamically assisted Schwinger mechanism.

Within the perturbation theory in the Furry picture (see Sec.~\ref{sec2a}), one can explicitly evaluate $N_{\rm LLL}$ as Eq.~(\ref{eq-34}).  Therefore, we have
\begin{align}
	Q_5 &= Q_5({\mathcal E}=0) \nonumber\\
	&\quad\times\! \Biggl[ 1 + \frac{2\pi}{T} \Biggl\{ \frac{1}{2} \frac{ m^2}{e\bar{E}}  \frac{\tilde{\mathcal E}(0)}{\bar{E}} + \frac{1}{4} \left( \frac{ m^2}{e\bar{E}} \right)^2 \int_0^{\infty}\!\! {\rm d}\omega \nonumber\\
	&\quad \quad \times\! \left| \frac{\tilde{\mathcal E}(\omega)}{\bar{E}} {}_1\tilde{F}_1 \left( \!1\!-\! \frac{\rm i}{2} \frac{m^2}{e\bar{E}}; 2; \frac{\rm i}{2} \frac{\omega^2}{e\bar{E}}\! \right)\! \right|^2 \!\Biggl\} \Biggl] ,  \label{eq---53}
\end{align}
where
\begin{align}
	Q_5({\mathcal E}=0) = VT \times \frac{e\bar{E}e\bar{B}}{2\pi^2} \exp\left[ -\pi \frac{ m^2}{e\bar{E}} \right] \label{eq---54}
\end{align}
is chirality production without any perturbations \cite{cop18, fuk10, war12}.  $Q_5 ({\mathcal E}=0)$ is exponentially suppressed by the mass $m$, and hence chirality production for massive particles is usually negligible.  However, the dynamical assistance by a perturbation,
\begin{align}
	\frac{Q_5({\mathcal E})}{Q_5({\mathcal E}=0)} -1 > 0,
\end{align}
is always positive and, as we shall demonstrate below, can enhance chirality production by many orders of the magnitude. Hence, a sizable amount of chirality can be produced even for massive particles.  Note that in the massless limit $m \to 0$, the dynamical assistance goes away because the Schwinger mechanism dominates the production, and we have
\begin{align}
	Q_5 \xrightarrow{m\to 0}{} Q_5 ({\mathcal E}=0) = VT \times \frac{e\bar{E}e\bar{B}}{2\pi^2}, \label{eq--56}
\end{align}
which is consistent with the Adler-Bell-Jackiw (ABJ) anomaly relation for $m=0$: $\partial_t (Q_5/V) = e\bar{E}e\bar{B}/{2\pi^2}$ \cite{adl69, bel69}.  Equation~(\ref{eq--56}) means that the dynamical assistance is not important for chirality production for light particles whose mass is sufficiently lighter than the electric field strength $e\bar{E}$.

\subsection{The dynamical assistance to chirality production} \label{sec3b}

\begin{figure}[!t]
\begin{center}
\includegraphics[width=0.49\textwidth]{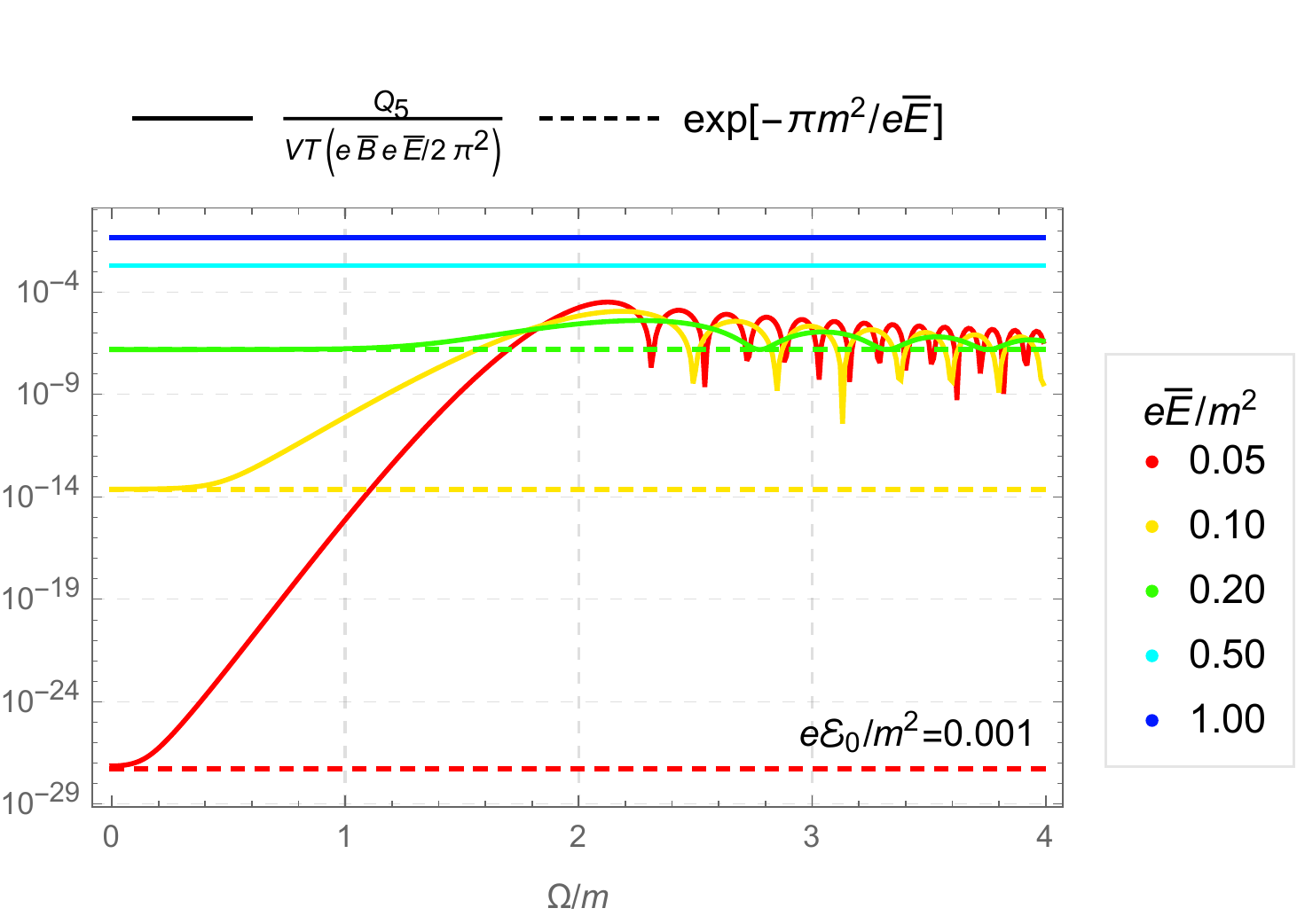}
\caption{\label{fig-5} Chirality production $Q_5$ (\ref{eq---53}) scaled by the anomaly factor $e\bar{E}e\bar{B}/2\pi^2$ for the monochromatic perturbation (\ref{eq-46}) as a function of the frequency $\Omega/m$.  Different colors distinguish electric field strengths $e\bar{E}/m^2 = 0.05, 0.10, 0.20, 0.50$, and $1.00$.  The dashed lines represent the mass suppression without perturbations (\ref{eq---54}).  The strength of the perturbative electric field is fixed as $e{\mathcal E}_0/m^2=0.001$.  Note that the result is independent of the magnetic field strength $e\bar{B}/m^2$.  }
\end{center}
\end{figure}

We evaluate the formula (\ref{eq---53}) with the monochromatic perturbation (\ref{eq-46}) to quantitatively discuss how the dynamical assistance modifies chirality production.  We display the result in Fig.~\ref{fig-5}.

The dynamical assistance by a perturbation can significantly enhance chirality production for massive particles.  The size of the enhancement strongly depends on the frequency of a perturbation $\Omega$.  For frequencies below the mass gap $\Omega \ll 2m$, the enhancement is not significant and chirality production is exponentially suppressed by the mass as Eq.~(\ref{eq---54}).  The enhancement becomes the maximum at around the energy threshold for the lowest Landau level $\Omega \sim 2m$.  This is because the dynamically assisted Schwinger mechanism takes place, i.e., the threshold energy is effectively reduced by the energy supply by a perturbation and thus the Schwinger mechanism for the lowest Landau level is enhanced, which results in the enhancement of chirality production.  Above the energy threshold $\Omega \gtrsim 2m$, the enhancement decays slowly and shows an oscillating dependence on $\Omega$, which can be understood as an analog of the Franz-Keldysh oscillation \cite{tah63, cal63, fk, fk2}.  Note that the enhancement increases quadratically with $e{\mathcal E}_0$ and that the $\Omega$-dependence is unchanged with $e{\mathcal E}_0$ unless $e{\mathcal E}_0$ becomes so large that the formula (\ref{eq---53}), which only takes into account the lowest correction from a perturbation, becomes invalid.

The enhancement is not significant for light particles, as we discussed below Eq.~(\ref{eq--56}).  This is because chirality production is free from the exponential suppression for light particles, and hence the dynamical assistance, which is always suppressed by powers of the mass, can only give a secondary contribution.  Note that the insensitivity to the time-dependence of a field for chirality production for light particles and the decrease of that for massive particles for a slower field configuration are consistent with Ref.~\cite{amb83}, in which chirality production for massless and massive particles by a pulsed electric field with a sudden switching on/off was discussed.

\section{Summary and discussion} \label{sec4}

We have studied particle production from the vacuum by a slow strong parallel electromagnetic field superimposed by a fast weak perturbation.  We have generalized the perturbation theory in the Furry picture  \cite{fur51, fra81, fra91, gre17, gre18, gre19, fk, fk2} to include a magnetic component in the slow strong field configuration and derived an analytical formula for the production number.  Based on the formula, we have analytically discussed the particle production, focusing on the interplay between the Schwinger mechanism and one-photon pair production and the dynamical assistance between the two production mechanisms.  In particular, we have shown that (i) the strength of the strong electric field and the frequency of a perturbation control the interplay, and the Schwinger mechanism (one-photon pair production) dominates the particle production when the electric field strength is large (small) and the frequency is small (large); (ii) the two production mechanisms occur at the same time and assist each other at intermediate values of the electric field strength and the frequency, and the particle production is significantly enhanced by many orders of the magnitude (i.e., the dynamically assisted Schwinger mechanism \cite{sch08, piz09, dun09, mon10a, mon10b}); (iii) the effect of a strong magnetic field is to discretize the energy level through the Landau quantization, which enhances the production proportionally to the magnetic field strength because of the enhancement of the phase-space, and results in sharp threshold behaviors in the production number when the magnetic field strength is comparable to or stronger than the electric one; and (iv) the lowest Landau approximation becomes valid for a very strong magnetic field compared to mass and the electric field strength, for which the production number is maximized at frequency around the energy threshold $\Omega \sim 2m$ and exhibits oscillating dependence on $\Omega$ for $\Omega \gtrsim 2m$ (an analog of the Franz-Keldysh oscillation \cite{tah63, cal63, fk, fk2}).

We also have clarified how the dynamical assistance to the particle production affects chirality production.  We have explicitly evaluated the in-in vacuum expectation value of the chirality operator and shown that chirality production is determined solely by the lowest Landau level production.  Therefore, chirality production is enhanced because the lowest Landau level production is enhanced by the dynamical assistance.  We have written down an analytical formula for chirality production based on the production number formula within the perturbation theory in the Furry picture, and shown that (i) chirality production for light particles whose mass is lighter than the electric field strength is less affected by the dynamical assistance; (ii) chirality production for massive particles is significantly enhanced by many orders of the magnitude; and (iii) the enhancement becomes the largest at frequency $\Omega \sim 2m$, where the dynamical assistance to the lowest Landau level production is maximized.

An interesting application of our results is in heavy-ion collisions at RHIC and the LHC.  We have shown in the present paper that perturbations on top of a slow strong parallel electromagnetic field significantly affect particle and chirality production.  In heavy-ion collisions, the glasma \cite{low75, nus75, kov95a, kov95b, lap06} plays the role of a slow strong parallel (chromo-)electromagnetic field and a weak fast perturbation is naturally seeded by quark/gluon jets produced by initial hard collisions.  Therefore, we expect that particle and chirality production in heavy-ion collisions should be modified significantly by jets on top of the glasma.  For example, a sizable amount of chirality may be produced even for massive strange and charm quarks because of the dynamical assistance by jets.  This could leave experimental signatures such as charge asymmetry of heavy hadrons through the chiral magnetic effect.  Another example is that the enhancement of the particle production by jets may speed up the formation of the quark-gluon plasma in heavy-ion collisions, which could be interesting to the early thermalization/hydrodynamization puzzle \cite{hei05, boz11}.  We may also expect that propagation of jets should be modified significantly by the glasma because jets lose their energy via the particle production in the glasma.  This could give an additional contribution to the jet quenching and broadening phenomena.

Another interesting application is in intense laser experiments at, e.g., ELI and XCELS.  Experimentally, one may realize a parallel slow strong electromagnetic field configuration by colliding two counter propagating laser beams with different polarization.  For example, by colliding two laser beams described by gauge potentials $\bar{\bm A}_1 = (a/k) \sin (k (x^1 - x^0)) \times (0,1,0)$ and $\bar{\bm A}_2 = (a/k) \sin (k (-x^1 - x^0)) \times (0,\sin \phi, \cos \phi)$ and assuming that the laser beams are sufficiently slow $|k|\ll 1$ and that the particle production occurs at the region $|kx^1|\ll1$, one gets 
\begin{align}
	\bar{\bm E} \sim a \begin{pmatrix} 0 \\ 1+\sin \phi \\ \cos\phi \end{pmatrix},\ 
	\bar{\bm B} \sim \frac{\cos\phi}{1+\sin \phi} \bar{\bm E}.  
\end{align}
Thus, we have $\bar{\bm E} \parallel \bar{\bm B}$ and can control the relative size between the electric and magnetic field strengths by tuning the polarization angle $\phi$ (e.g., $\bar{E}=\bar{B}\neq0$ for $\phi=0$; $\bar{E}=0, \bar{B}\neq0$ for $\phi=-\pi/2$; and $\bar{E}\neq0, \bar{B}=0$ for $\phi=\pi/2$).  As shown in the present paper, the existence of a parallel strong magnetic field makes the particle production phenomenologically richer, e.g., it results in the non-trivial frequency dependence for a high-frequency perturbation and chirality production.  The sharp frequency dependence for strong $e\bar{B}$ may be detected by using techniques of modulation spectroscopy \cite{car69}, which is a well-established method to measure the Franz-Keldysh oscillation in semi-conductor physics.  Observing chirality production is not only new to laser physics, but also important to understand the chirality production mechanism in heavy-ion collisions and consequent observables, which are less understood in the heavy-ion community despite great experimental/theoretical efforts over the past decade.  Also, chirality production may induce anomalous transport phenomena such as the chiral magnetic effect even in laser systems.  This is a novel opportunity to study anomalous transport phenomena and may open up a new connection between laser physics and the other areas of physics since anomalous transport phenomena have been attracting attention from broad areas of physics with different energy scales, such as Weyl/Dirac semi-metals, heavy-ion collisions, neutron stars, and supernovae.

For future work, it is interesting to extend our lowest order formula (\ref{eq64}) by including higher order $e^k$ ($k \geq 2$) corrections.  This enables us to discuss the dynamical assistance not only by one-photon pair production ($\gamma + \bar{E},\bar{B} \to e^+e^-$), but also by multi-photon pair production ($n \gamma + \bar{E},\bar{B} \to e^+e^-$) and multi-pair production by a photon ($\gamma + \bar{E},\bar{B} \to n\; e^+e^-$), which are parametrically suppressed by the strength of a perturbation as $(e{\mathcal E}/m^2)^{2n}$ but become important if $e{\mathcal E}$ becomes strong.  Such higher order processes may, for example, induce additional peak structures in the production number as a function of the frequency of a perturbation \cite{fk2, hei10, gui10, koh14, aka14}.  Another possible extension is to consider a perturbation with various polarizations.  Not only the production number/distribution, but also spin dynamics is modified by polarization, which results in non-trivial spin-accumulation and/or generation of a spin-current \cite{mat19, fk2, ant15, koh19}.  Non-trivial spin dependence/observables imply a modification to chirality production and, therefore, may induce novel chirality-dependent observables.  It is also important to understand polarization effects to discuss heavy-ion collisions since jets in heavy-ion collisions are not necessarily longitudinal with respect to the glasma.  It is very straightforward to extend our formalism to such a general perturbation with different polarization as well as, for example, with spatial dependence, which has been discussed within the worldline instanton formalism for a slowly varying case \cite{sch16, cop16} and within a numerical method \cite{ale17, ale18}.  The last direction that we would like to mention is to consider different types of perturbations.  In the present paper, we concentrated on a perturbation by a weak fast electromagnetic field (or a dynamical photon).  In principle, any kinds of perturbations (or, generally, external forces) may dynamically assist the Schwinger mechanism and vice versa as long as they supply energy to the vacuum.  Such a consideration has been done recently for vibrating plates (or the dynamical Casimir effect) \cite{tay20}.  A strong magnetic field drastically changes the dimensionality of the system because of the Landau quantization.  Hence, the dynamical assistance by geometric perturbations such as vibration of plates and gravitational fields should be modified.

\section*{Acknowledgment}
The author would like to thank Gerald~V.~Dunne, Kenji~Fukushima, Koichi~Hattori, and Xu-Guang~Huang for fruitful discussions.  This work was partially supported by National Natural Science Foundation in China (NSFC) under Grant No.~11847206.

\end{document}